\documentclass[letterpaper,journal]{IEEEtran}
\usepackage{amsmath,amsfonts}
\usepackage{algorithmic}
\usepackage{algorithm}
\usepackage{array}
\usepackage{subcaption}
\usepackage{textcomp}
\usepackage{stfloats}
\usepackage{url}
\usepackage{verbatim}
\usepackage{graphicx}
\usepackage{cite}
\usepackage{bm}
\usepackage{cancel}
\usepackage[colorlinks=true,linkcolor=blue,citecolor=blue,urlcolor=blue]{hyperref}
\hypersetup{pdftitle={Temporal Regression-Based Model-Free Sensorless Control of Permanent Magnet Synchronous Motor},
pdfsubject={Theory draft for IEEE Transactions on Industrial Electronics},
pdfkeywords={SPMSM, sensorless control, temporal regression, parameter independence}}
\newcommand{\dd}{\mathop{}\!\mathrm{d}}
\newcommand{\T}{\mathsf{T}}
\newcommand{\R}{\mathbb{R}}
\DeclareMathOperator{\col}{col}

\DeclareMathOperator{\tr}{tr}
\DeclareMathOperator*{\argmin}{arg\,min}

\begin{document}
\title{Temporal Regression-Based Model-Free Sensorless Control of Permanent Magnet Synchronous Motor}
\author{
Fobao Zhou, \emph{Student Member, IEEE}, Yang Shen, \emph{Student Member, IEEE}, Xueyan Wang, \\ Zhenxiao Yin, \emph{Member, IEEE}, Yujia Zhang, Yuanfeng Qu, and Hang Zhao$^\ast$, \emph{Member, IEEE}

\thanks{This work is supported by the National Natural Science Foundation of China (No. 52407066), Guangdong Science and Technology Program (2025A0505020029), and Youth S\&T Talent Support Program of Guangdong Provincial Association for Science and Technology (SKXRC2025464). (\textit{Corresponding author: Hang Zhao.)}}
\thanks{The authors are with the Robotics and Autonomous Systems Thrust, The
Hong Kong University of Science and Technology (Guangzhou), Guangzhou 511453, China
(e-mail: hangzhao@hkust-gz.edu.cn).}}
\markboth{Journal of \LaTeX\ Class Files}%
{Temporal Regression-Based Model-Free Sensorless Control of Permanent Magnet Synchronous Motor}
\maketitle 
\begin{abstract}
To address the widespread sensitivity of surface-mounted permanent magnet synchronous motor (SPMSM) sensorless control to motor parameters, this paper proposes a temporal regression-based model-free sensorless control (TFC) method. First, voltage integrals and current increments over consecutive short intervals are stacked to construct a finite window regression, in which the unknown stator inductance appears as a common scalar coefficient. Second, a projection operator constructed from the stacked current increments eliminates the inductance term, and a least-squares formulation is developed to reconstruct the rotor flux vector. Meanwhile, the analysis of the projected regression and current-flux geometry establishes a two-dimensional direction vector whose components share a common amplitude containing the stator resistance and flux linkage. This amplitude cancels during position extraction. By setting the resistance reference to zero, the proposed method estimates the position without specifying the stator resistance, inductance, or flux linkage. Finally, experimental results verify the effectiveness of the proposed TFC method.

\end{abstract}


\section{Introduction}
\label{sec:introduction}
\IEEEPARstart{P}{ermanent} magnet synchronous motors (PMSMs) are widely used in high performance electric drives, including electric vehicles, industrial servo systems, and robotics, owing to their high efficiency, high power density, compact structure, and favorable dynamic performance. High performance vector control requires accurate rotor position information. However, mechanical position sensors increase cost and size and can compromise reliability in harsh environments, motivating extensive research into position sensorless control \cite{8917922}. 

At medium and high speeds, sensorless control typically relies on the fundamental-frequency model to estimate the back electromotive force (EMF) or flux linkage and subsequently recover the rotor position. Representative approaches include kalman filter (KF) \cite{10946691,11197560}, sliding-mode observer (SMO) \cite{10510487,10820364}, extended state observer (ESO) \cite{10551501,10740488}, and flux observer \cite{11414226, 11099531}. Despite their different structures, these methods generally rely on the motor voltage equations and require the stator resistance, inductance, or flux linkage. However, these parameters vary with temperature and magnetic saturation under different operating conditions, and the resulting mismatches introduce back-EMF or flux estimation errors that degrade rotor position accuracy \cite{11270893}. Therefore, reducing this parameter dependence is essential for improving the robustness of sensorless control.

To address parameter sensitivity, considerable effort has been devoted to robust observer design or parameter-robust structures. In \cite{wang2025eso}, an ESO-based robust hybrid flux observer estimates and compensates for active flux errors caused by parameter mismatches, improving robustness to inductance and flux variations. In \cite{wang2026hybrid}, an improved flux observer incorporates a compensation mechanism to reduce the influence of stator resistance mismatch on position estimation.  In \cite{cheng2026terminal}, an adaptive full-order terminal SMO decouples the mutual inductance from the estimated active flux in speed observation, improving robustness to inductance variations. In \cite{tan2025sliding}, an SMO-based predictive observer combines an improved observer structure with parameter identification to enhance position estimation under parameter variations. These approaches nevertheless retain the motor model and attenuate the effects of parameter mismatches through robust design or error compensation, rather than eliminating parameter dependence. Recently, an ultralocal model-based sensorless method has been developed in \cite{11204721} to reduce reliance on the conventional motor model. However, it still requires the design of additional model parameters, and residual sensitivity to substantial inductance mismatches remains a concern. Overall, existing robust observers and parameter-robust controllers not only rely on motor parameters but also introduce additional observer or compensation gains, increasing the tuning effort. Therefore, sensorless methods that avoid both physical parameter dependence and complicated parameter tuning are of significant research value.

An alternative approach is to identify parameters online and update the position observer accordingly. In \cite{inoue2011identification}, online resistance identification improves position estimation at low speeds, although the inductance must still be determined beforehand. In \cite{yao2020adaptive}, adaptive full-state feedback control enables online estimation of the resistance, inductance, and flux linkage, and the estimates are used to improve position estimation accuracy. In \cite{wang2025virtual}, virtual back-EMF injection constructs a full-rank identification model for joint estimation of these three parameters. In \cite{hu2025deadbeat}, parameter mismatch detection and online inductance identification are incorporated into robust predictive sensorless control to compensate for steady-state parameter errors. Under sensorless operation, however, the motor parameters, back-EMF, and rotor position are simultaneously unknown. Parameter identification is therefore subject to rank deficiency, parameter coupling, and interactions between position estimation errors and parameter estimation errors. Although online identification can mitigate parameter mismatches, it also introduces additional computational, tuning, and convergence considerations \cite{10458351,10750528}.

To eliminate the dependence of surface-mounted PMSM (SPMSM) sensorless position estimation on motor parameters while avoiding the complexity of online parameter identification, elaborate error compensation, and extensive observer tuning, this paper proposes a temporal regression-based model-free sensorless control (TFC) method. 
The proposed method exploits temporal relationships in measured voltage and current data to recover rotor position without requiring motor parameter inputs. Its central feature is the separation of position information from parameter dependent components through temporal projection and the current-flux geometry of the SPMSM. This formulation eliminates the need for online motor parameter identification and reduces the complexity associated with mismatch compensation and observer tuning, providing a direct approach to parameter independent sensorless position estimation. This paper makes several key contributions and innovations:

\begin{enumerate}
\item A model-free sensorless control method is developed to reconstruct rotor position from voltage and current data without requiring stator resistance, inductance, or flux linkage as inputs or identifying them online.
\item A temporal projection framework is introduced to eliminate the unknown inductance directly from a finite window data regression, exploiting its shared coefficient across consecutive intervals.
\item The current-flux geometry is exploited to incorporate the effects of resistance and flux linkage into a common amplitude factor of the
direction vector, and this factor cancels during angle extraction.
\item The resulting estimation structure primarily involves two tuning parameters, the window size and the interval duration, which simplifies implementation and tuning.
\end{enumerate}

\section{SPMSM Model and Current-Flux Geometry}
\label{sec:model}

\subsection{SPMSM Model}
\label{subsec:model}
The SPMSM under study is modeled as a two-dimensional vector system in the stationary $\alpha\beta$ reference frame. Let $\bm u$, $\bm i$, and $\bm\lambda$ denote the stator voltage vector, current vector, and flux vector, respectively, defined as:
\begin{equation}
\bm u=\begin{bmatrix}u_\alpha\\u_\beta\end{bmatrix},\quad
\bm i=\begin{bmatrix}i_\alpha\\i_\beta\end{bmatrix},\quad
\bm\lambda=\begin{bmatrix}\lambda_\alpha\\\lambda_\beta\end{bmatrix}
\label{eq:electrical_vectors}
\end{equation}

Let $L_s>0$ and $R_s>0$ denote the actual stator inductance and stator resistance, respectively. The SPMSM model is:
\begin{equation}
\bm\lambda=L_s\bm i+\bm x,\quad
\dot{\bm\lambda}=\bm u-R_s\bm i
\label{eq:spmsm_model}
\end{equation}
where $\bm x\in\R^2$ is the flux vector.

Let $\theta$ denote the rotor electrical angle, and let $\psi_f>0$ denote the flux linkage. The flux vector is:
\begin{equation}
\bm x=\psi_f\bm c(\theta),\quad
\bm c(\theta)=
\begin{bmatrix}\cos\theta\\\sin\theta\end{bmatrix}
\label{eq:flux_direction}
\end{equation}
where $\bm c(\theta)$ is the unit vector along the rotor $d$-axis. Consequently, the flux model becomes
$\bm\lambda=L_s\bm i+\psi_f\bm c(\theta)$.
The proposed estimator uses these electromagnetic relationships to construct a regression from measured data.

\subsection{Current-Flux Orthogonality}
\label{subsec:orthogonality}
Define the counterclockwise $90^\circ$ rotation matrix $\bm J$ and the two-dimensional identity matrix $\bm I_2$ by:
\begin{equation}
\bm J=\begin{bmatrix}0&-1\\1&0\end{bmatrix},\quad
\bm J^\T=-\bm J,\quad \bm J^2=-\bm I_2
\label{eq:J}
\end{equation}

The vector $\bm J\bm c(\theta)$ is the unit vector along the rotor $q$-axis. Let $i_d$ and $i_q$ denote the stator current components along the actual rotor $d$- and $q$-axis, respectively. The stationary-frame current then satisfies:
\begin{equation}
\bm i=i_d\bm c(\theta)+i_q\bm J\bm c(\theta)
\label{eq:dq_current}
\end{equation}

In field-oriented control with $i_d=0$, \eqref{eq:dq_current} satisfies the following conditions:
\begin{equation}
\bm i=i_q\bm J\bm c(\theta)
=\frac{i_q}{\psi_f}\bm J\bm x,\quad
\bm x^\T\bm i=0
\label{eq:current_flux_geometry}
\end{equation}

Equation \eqref{eq:current_flux_geometry} shows that the stator current is orthogonal to the flux vector. This geometric constraint provides the direction information required for rotor position estimation after the inductance contribution has been removed.

\section{Temporal Regression-Based Model-Free Sensorless Control Method}
\label{sec:method}
This section presents the proposed TFC method in detail. Voltage and current measurements over consecutive short intervals are stacked into a regression whose common inductance term is removed by projection, and the rotor position is then recovered from the projected data.

\subsection{Finite Window Representation}
\label{subsec:rotation}
For a constant flux linkage within the data window, differentiation of \eqref{eq:flux_direction} gives:
\begin{equation}
\dot{\bm x}=\omega\bm J\bm x,\quad
\omega=\dot{\theta}
\label{eq:flux_dynamics}
\end{equation}
where $\omega$ is the rotor electrical angular velocity. The planar rotation matrix $\bm C(\varphi)$ is defined as:
\begin{equation}
\bm C(\varphi)=
\begin{bmatrix}
\cos\varphi&-\sin\varphi\\
\sin\varphi& \cos\varphi
\end{bmatrix}
\label{eq:C}
\end{equation}
where $\varphi$ is the position. This matrix satisfies
$\bm C^\T(\varphi)\bm C(\varphi)=\bm I_2$ and
$\bm J\bm C(\varphi)=\bm C(\varphi)\bm J$.

Consider $N\geq2$ consecutive intervals with boundary times
$t_0<t_1<\cdots<t_N$, where $N$ is the number of intervals and $t_N$ is the current window endpoint. Let $\bm x_N=\bm x(t_N)$ denote the flux vector to be reconstructed. For the boundary index $j=0,\ldots,N$, define the relative electrical angle:
\begin{equation}
\varphi_j=\theta(t_j)-\theta(t_N)
=-\int_{t_j}^{t_N}\omega(\tau)\dd\tau
\label{eq:relative_angle}
\end{equation}

The flux vector at each boundary can be expressed relative to the current endpoint as:
\begin{equation}
\bm x(t_j)=\bm C(\varphi_j)\bm x_N
\label{eq:rotated_flux}
\end{equation}

Equation \eqref{eq:rotated_flux} is exact for a constant $\psi_f$, including when speed varies, provided that the relative angles are known. For a sufficiently short window with slowly varying speed, the relative angles can be approximated by:
\begin{equation}
\varphi_j\simeq\omega(t_j-t_N)
\label{eq:local_speed}
\end{equation}

In implementation, $\omega$ is replaced by the estimated electrical speed $\hat\omega$, so that $\varphi_j\simeq\hat\omega(t_j-t_N)$.

\subsection{Short Interval Integral Regression}
\label{subsec:regression}
To analyze resistance dependence, introduce an algorithmic resistance $\hat R_s$ and its mismatch:
\begin{equation}
\widetilde R_s=\hat R_s-R_s
\label{eq:Rrror}
\end{equation}
where $\widetilde R_s$ is the resistance mismatch.

For interval $j=1,\ldots,N$, define the voltage integral $\bm z_j$, current integral $\bm h_j$, and current increment $\bm d_j$ as:
\begin{align}
\bm z_j&=\int_{t_{j-1}}^{t_j}
       [\bm u(\tau)-\hat R_s\bm i(\tau)]\dd\tau,
\label{eq:z}\\
\bm h_j&=\int_{t_{j-1}}^{t_j}\bm i(\tau)\dd\tau,\quad
\bm d_j=\bm i(t_j)-\bm i(t_{j-1})
\label{eq:hd}
\end{align}
where $\bm z_j$, $\bm h_j$ and $\bm d_j$ are constructed from voltage and current samples. With $\hat R_s=0$, $\bm z_j$ is simply the voltage integral, and $\bm h_j$ is needed only for the resistance analysis.

Integrating \eqref{eq:spmsm_model} over each interval yields:
\begin{equation}
\bm z_j=L_s\bm d_j+\Delta\bm x_j-\widetilde R_s\bm h_j
\label{eq:interval_regression}
\end{equation}
where $\Delta\bm x_j=\bm x(t_j)-\bm x(t_{j-1})$ is the flux increment. Using \eqref{eq:rotated_flux}, the interval rotation matrix $\bm A_j$ is defined as:
\begin{equation}
\bm A_j=\bm C(\varphi_j)-\bm C(\varphi_{j-1}),\quad
\Delta\bm x_j=\bm A_j\bm x_N
\label{eq:Aj}
\end{equation}

Substituting \eqref{eq:Aj} into \eqref{eq:interval_regression} yields the regression model:
\begin{equation}
\bm z_j=L_s\bm d_j+\bm A_j\bm x_N-\widetilde R_s\bm h_j
\label{eq:interval_linear}
\end{equation}

The unknown flux vector at the window endpoint enters linearly, while the inductance multiplies a measured current increment. Finite interval integration avoids current differentiation and requires no initial stator flux value.

\subsection{Temporal Projection and Inductance Elimination}
\label{subsec:projection}
Let $\col(\cdot)$ denote vertical concatenation of vectors or matrices. Stack the interval quantities as:
\begin{equation}
\begin{aligned}
\bm Z&=\col(\bm z_1,\ldots,\bm z_N)\in\R^{2N},\\
\bm D&=\col(\bm d_1,\ldots,\bm d_N)\in\R^{2N},\\
\bm H&=\col(\bm h_1,\ldots,\bm h_N)\in\R^{2N},\\
\mathcal A&=\col(\bm A_1,\ldots,\bm A_N)\in\R^{2N\times2}
\end{aligned}
\label{eq:stack}
\end{equation}
where $\bm Z$, $\bm D$, and $\bm H$ collect the voltage integrals, current increments, and current integrals of the $N$ intervals, respectively, and $\mathcal A$ collects the corresponding interval rotation matrices.

The extended regression becomes:
\begin{equation}
\bm Z=\bm D L_s+\mathcal A\bm x_N-\widetilde R_s\bm H
\label{eq:extended}
\end{equation}

All intervals share the same unknown scalar $L_s$. Therefore, its complete contribution lies in the one-dimensional subspace spanned by the measured vector $\bm D$.

For a window satisfying $\bm D^\T\bm D>0$, define the temporal projection matrix:
\begin{equation}
\bm P_L=\bm I_{2N}
-\frac{\bm D\bm D^\T}{\bm D^\T\bm D}
\label{eq:PL}
\end{equation}
where $\bm I_{2N}$ is the identity matrix. The matrix $\bm P_L$ projects onto the orthogonal complement of $\bm D$ and satisfies:
\begin{equation}
\bm P_L^\T=\bm P_L,\quad
\bm P_L^2=\bm P_L,\quad
\bm P_L\bm D=\bm0
\label{eq:projection_properties}
\end{equation}

Left multiplying \eqref{eq:extended} by $\bm P_L$ gives:
\begin{equation}
\bm P_L\bm Z=\bm P_L\mathcal A\bm x_N
-\widetilde R_s\bm P_L\bm H
\label{eq:projected}
\end{equation}

As shown in \eqref{eq:projected}, the unknown inductance term is eliminated by the temporal projection, yielding an inductance-free regression equation.

For the actual resistance, where $\widetilde R_s=0$, this regression is solved by the least-squares method:
\begin{equation}
\hat{\bm x}_N = \argmin_{\bm\xi\in\R^2}
\left \| \bm P_L\bm Z-\bm P_L\mathcal A\bm\xi\right\|_2^2
\label{eq:LS}
\end{equation}
where $\hat{\bm x}_N$ is the estimated flux vector at the current window endpoint and $\bm\xi$ is a candidate flux vector. Define the Gram matrix $\bm G$ and the data direction vector $\bm Y$ by:
\begin{equation}
\bm G=\mathcal A^\T\bm P_L\mathcal A,\quad
\bm Y=\mathcal A^\T\bm P_L\bm Z
\label{eq:GY}
\end{equation}

From \eqref{eq:projected} and \eqref{eq:GY}, the normal equation of \eqref{eq:LS} and its minimum norm solution are as follows:
\begin{equation}
\bm G\hat{\bm x}_N=\bm Y,\quad
\hat{\bm x}_N=\bm G^\dagger\bm Y
\label{eq:minimum_norm}
\end{equation}
where $(\cdot)^\dagger$ denotes the Moore-Penrose pseudoinverse.

Therefore, the temporal projection constructs an inductance-free regression model and provides an explicit solution for the estimated flux vector. The rotor position can then be obtained from the direction of this estimated flux vector.

\subsection{Flux Direction Reconstruction and Angle Extraction}
\label{subsec:angle}
For matched resistance, \eqref{eq:projected} and \eqref{eq:GY} give $\bm Y=\bm G\bm x_N$. The rotational excitation factor $\chi$ is defined as:
\begin{equation}
\chi=\frac{1}{2}\tr(\mathcal A^\T\mathcal A)
\label{eq:chi_general}
\end{equation}

Under the operating conditions analyzed later, the projected data vector is aligned with the endpoint rotor flux direction:
\begin{equation}
\bm G\bm x_N=\chi\bm x_N,\quad
\bm Y=\chi\psi_f
\begin{bmatrix}
\cos\theta(t_N)\\ \sin\theta(t_N)
\end{bmatrix}
\label{eq:ideal_direction}
\end{equation}

These conditions include equally spaced intervals, locally constant speed, zero
$d$-axis current, constant $q$-axis current, and nonzero rotational excitation;
the proof is given in Section~\ref{sec:analysis}. Thus, the components of $\bm Y$
contain the desired position and share the same positive amplitude $\chi\psi_f$.

Let $Y_\alpha$ and $Y_\beta$ denote the stationary-axis components of $\bm Y=[Y_\alpha,Y_\beta]^\T$. The rotor position estimate at the window endpoint is:
\begin{equation}
\hat\theta(t_N)=
\tan^{-1}\!\left(\frac{Y_\beta}{Y_\alpha}\right)
\label{eq:angle}
\end{equation}
where $\hat\theta(t_N)$ is the estimated rotor position at the window endpoint. The unknown flux linkage cancels as a common amplitude; it is not needed to normalize the vector. Under resistance mismatch, Section~\ref{subsec:resistance} establishes the corresponding positive amplitude condition.

To obtain an explicit expression for $\bm Y$, substitute the projection matrix in \eqref{eq:PL} into its definition in \eqref{eq:GY}. Distributing the matrix products gives:
\begin{equation}
\begin{aligned}
\bm Y
&=\mathcal A^\T\left(\bm I_{2N}-\frac{\bm D\bm D^\T}{\bm D^\T\bm D}\right)\bm Z\\
&=\mathcal A^\T\bm Z
-\frac{\mathcal A^\T\bm D\bm D^\T\bm Z}{\bm D^\T\bm D}.
\end{aligned}
\end{equation}

Since $\bm D^\T\bm D$ is a nonzero scalar, associativity allows the numerator of the second term to be evaluated as $(\mathcal A^\T\bm D)(\bm D^\T\bm Z)$, where $\mathcal A^\T\bm D$ is a two-dimensional vector and $\bm D^\T\bm Z$ is a scalar. The vector can be evaluated without explicitly storing the $2N\times2N$ projection matrix:
\begin{equation}
\bm Y=\mathcal A^\T\bm Z
-\frac{(\mathcal A^\T\bm D)(\bm D^\T\bm Z)}
{\bm D^\T\bm D}
\label{eq:Yfficient}
\end{equation}

\begin{figure*}[!t]
    \centerline{\includegraphics[width=46em]{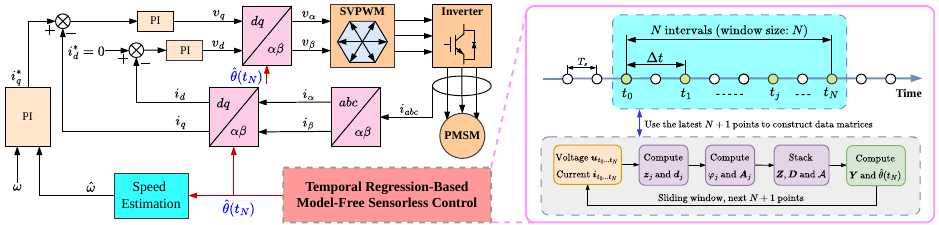}}
    \caption{Block diagram of the proposed TFC method for sensorless control.}
    \label{Structure diagram}
\end{figure*}

Each update integrates new data, advances the window, forms relative rotation matrices, and evaluates \eqref{eq:angle} and \eqref{eq:Yfficient}, from which the estimated rotor position at the window endpoint is obtained. Fig.~\ref{Structure diagram} shows the block diagram of the method. The implementation is summarized in Algorithm~\ref{alg:temporal_projection}.

\section{Theoretical Analysis and Parameter Independence}
\label{sec:analysis}
This section theoretically analyzes the proposed TFC method. It reveals how the rotor flux direction is preserved after the temporal projection and the associated matrix mapping, and how the proposed method avoids the need for resistance and flux linkage information.

\subsection{Properties of the Temporal Rotation Matrix}
\label{subsec:rotation_properties}
Consider equally spaced interval boundaries with interval duration $\Delta t=t_j-t_{j-1}>0$. Under constant electrical speed, define the electrical angle increment over each interval as $\delta=\omega\Delta t$. The relative angles satisfy $\varphi_j-\varphi_{j-1}=\delta$, and expansion of \eqref{eq:Aj} gives:
\begin{align}
\bm A_j^\T\bm A_j
&=2\bm I_2-\bm C(\delta)-\bm C(-\delta)\nonumber\\
&=2(1-\cos\delta)\bm I_2\nonumber\\
&=4\sin^2(\delta/2)\bm I_2
\label{eq:Aj_gram}
\end{align}

Therefore,
\begin{equation}
\mathcal A^\T\mathcal A=\chi\bm I_2,\quad
\chi=4N\sin^2\!\left(\frac{\omega\Delta t}{2}\right)
\label{eq:ATA}
\end{equation}

The two columns of $\mathcal A$ are orthogonal and have equal squared norm $\chi$. This isotropic Gram matrix allows temporal projection to be related to a spatial direction.

When $\chi>0$, the time window contains nonzero flux rotation information, which can be further used for position direction recovery. Moreover, for fixed $N$ and $\Delta t$, the direction signal weakens as the speed decreases.

\begin{algorithm}[!t]
\caption{ Proposed Model-Free Sensorless Control}
\label{alg:temporal_projection}
\begin{algorithmic}[1]
\STATE \textbf{Control input:} window size $N$, interval duration $\Delta t$
\STATE \textbf{Optional:} algorithmic resistance $\hat R_s$ (default: $0$)
\STATE \textbf{Output:} position estimate $\hat\theta(t_N)$
\REPEAT
    \STATE Measure stator voltage $\bm u$ and current $\bm i$.
    \STATE Update the window of $N$ consecutive intervals.
    \STATE Compute $\bm z_j$ and $\bm d_j$ by \eqref{eq:z} and \eqref{eq:hd}.
    \STATE Compute $\varphi_j$ by \eqref{eq:local_speed} using $\hat\omega$.
    \STATE Form $\bm A_j$ by \eqref{eq:Aj}.
    \STATE Stack $\bm Z$, $\bm D$, and $\mathcal A$ by \eqref{eq:stack}.
    \STATE Compute the direction vector $\bm Y$ by \eqref{eq:Yfficient}.
    \STATE Recover $\hat\theta(t_N)$ from $\bm Y$ by \eqref{eq:angle}.
    \STATE Estimate speed $\hat\omega$ from $\hat\theta(t_N)$.
\UNTIL{the motor is stopped}
\end{algorithmic}
\end{algorithm}

\subsection{Geometry of Current Increments}
\label{subsec:current_data}
Assume $i_d=0$ and a constant nonzero $q$-axis current $i_q$ throughout the window. By \eqref{eq:current_flux_geometry}, \eqref{eq:rotated_flux}, and the commutation of $\bm J$ and $\bm C$, we have:
\begin{equation}
\bm i(t_j)=\bm C(\varphi_j)\bm q,\quad
\bm q=\frac{i_q}{\psi_f}\bm J\bm x_N
\label{eq:q}
\end{equation}
where $\bm q\in\R^2$ is the current vector expressed at the window endpoint. In particular, $\bm q=\bm i(t_N)$, $\bm q^\T\bm x_N=0$, and $\bm q^\T\bm q=i_q^2$. The vector is parallel to the endpoint $q$-axis, with its orientation determined by the sign of $i_q$.

Taking adjacent differences yields:
\begin{equation}
\bm d_j=\bm A_j\bm q,\quad
\bm D=\mathcal A\bm q
\label{eq:D_Aq}
\end{equation}

Combining \eqref{eq:D_Aq} with \eqref{eq:ATA} gives:
\begin{equation}
\mathcal A^\T\bm D=\chi\bm q,\quad
\bm D^\T\bm D=\chi\bm q^\T\bm q=\chi i_q^2
\label{eq:currentnergy}
\end{equation}

Equation \eqref{eq:currentnergy} establishes an important geometric relation between the temporally extended current data $\bm D$ and the flux direction at the window endpoint.

\subsection{Preservation of the Rotor Flux Direction}
\label{subsec:preservation}
Substituting \eqref{eq:PL} into the definition of $\bm G$ gives:
\begin{equation}
\bm G=\mathcal A^\T\mathcal A
-\frac{(\mathcal A^\T\bm D)(\mathcal A^\T\bm D)^\T}
{\bm D^\T\bm D}
\label{eq:Gxpand}
\end{equation}

For $\chi>0$ and $i_q\neq0$, substituting \eqref{eq:ATA} and \eqref{eq:currentnergy} into \eqref{eq:Gxpand} reduces it to:
\begin{equation}
\bm G=\chi\left(\bm I_2-
\frac{\bm q\bm q^\T}{\bm q^\T\bm q}\right)
\label{eq:G_q}
\end{equation}

Since $\bm q$ is orthogonal to $\bm x_N$, it follows from the two-dimensional orthogonal basis identity that:
\begin{equation}
\bm I_2-\frac{\bm q\bm q^\T}{\bm q^\T\bm q}
=\frac{\bm x_N\bm x_N^\T}{\|\bm x_N\|_2^2}
\label{eq:orthogonal_basis}
\end{equation}

The projected Gram matrix therefore has the explicit form:
\begin{equation}
\bm G=\chi\frac{\bm x_N\bm x_N^\T}{\|\bm x_N\|_2^2},
\quad \bm G\bm x_N=\chi\bm x_N
\label{eq:G_flux}
\end{equation}

The analysis above constitutes the proof of \eqref{eq:ideal_direction}. Temporal projection retains the rotor $d$-axis component: $\bm G$ has eigenvalue $\chi$ along $\bm x_N$ and zero along $\bm J\bm x_N$. Although this rank-one matrix has no conventional inverse, the physical flux lies in its retained subspace. Its pseudoinverse is:
\begin{equation}
\bm G^\dagger=\frac{1}{\chi}
\frac{\bm x_N\bm x_N^\T}{\|\bm x_N\|_2^2}
\label{eq:G_pinv}
\end{equation}

For matched resistance, $\bm Y=\chi\bm x_N$, and hence
$\bm G^\dagger\bm Y=\bm x_N$. The minimum-norm least-squares solution therefore coincides with the physical flux under the stated geometry. Since $\chi>0$, $\bm Y=\chi\bm x_N$ directly establishes the correct orientation in \eqref{eq:angle}. Preserving the flux subspace is sufficient; full column rank of the projected regression is unnecessary.

\subsection{Resistance and Flux Independence}
\label{subsec:resistance}
Consider a constant resistance mismatch $\widetilde R_s$ over a window with $i_d=0$, constant $i_q\neq0$, and constant $\omega\neq0$. Combining \eqref{eq:current_flux_geometry} and \eqref{eq:flux_dynamics}, we have:
\begin{equation}
\bm i=\frac{i_q}{\psi_f\omega}\dot{\bm x}
\label{eq:i_flux_derivative}
\end{equation}

 Integrating \eqref{eq:i_flux_derivative} and using \eqref{eq:Aj} yields:
\begin{equation}
\bm h_j=\frac{i_q}{\psi_f\omega}\bm A_j\bm x_N,\quad
\bm H=\frac{i_q}{\psi_f\omega}\mathcal A\bm x_N
\label{eq:H}
\end{equation}


Substitution of \eqref{eq:H} into \eqref{eq:projected} and \eqref{eq:GY} yields:
\begin{align}
\bm Y
&=\bm G\bm x_N-\widetilde R_s\mathcal A^\T\bm P_L\bm H
\nonumber\\
&=\left(1-\frac{\widetilde R_s i_q}{\psi_f\omega}\right)
\bm G\bm x_N
=\chi\kappa_R\bm x_N
\label{eq:Y_resistance}
\end{align}
where the resistance scaling factor is:
\begin{equation}
\kappa_R=1-\frac{\widetilde R_s i_q}{\psi_f\omega}
\label{eq:kappa}
\end{equation}

Define the common amplitude $\Psi  =\chi\kappa_R\psi_f$. The direction-vector components are then:
\begin{equation}
Y_\alpha=\Psi\cos\hat{\theta}(t_N),\quad
Y_\beta=\Psi\sin\hat{\theta}(t_N)
\label{eq:common_amplitude}
\end{equation}

Both the flux linkage and the resistance mismatch enter through $\Psi$. Whenever $\chi>0$, $\psi_f>0$, and $\kappa_R>0$, this amplitude is positive and cancels in the ratio $Y_\beta/Y_\alpha$:
\begin{equation}
\hat\theta(t_N)=\tan^{-1}\!\left(
\frac{\cancel{\Psi}\sin \hat{\theta}(t_N)}{\cancel{\Psi}\cos\hat{\theta}(t_N)}\right)
\label{eq:angle_independence}
\end{equation}

From \eqref{eq:angle_independence}, the resistance mismatch rescales the flux term without changing its direction. A fully parameter-independent realization follows by choosing $\hat R_s=0$. In this case, we have:
\begin{equation}
\begin{aligned}
\bm z_j&=\int_{t_{j-1}}^{t_j}\bm u(\tau)\dd\tau,\\
\kappa_R&=1+\frac{R_s i_q}{\psi_f\omega},\\
\bm Y&=\chi\left(\psi_f+\frac{R_s i_q}{\omega}\right)
\bm c\bigl(\hat{\theta}(t_N)\bigr)
\end{aligned}
\label{eq:zero_R}
\end{equation}

Thus, \eqref{eq:Yfficient} and \eqref{eq:angle} can be evaluated from voltage and current data and estimated relative rotations without specifying $R_s$, $L_s$, or $\psi_f$. The actual resistance and flux linkage remain physical determinants of the signal amplitude, but their values are absent from the angle computation.

In summary, the proposed TFC method does not require the motor parameters in the rotor position estimation:
\begin{itemize}
\item The stator inductance $L_s$ is removed by temporal projection, because its contribution lies in the subspace spanned by the measured current increment vector.
\item The stator resistance $R_s$ is not used as an input parameter. With $\hat R_s=0$, its effect only appears as a scalar factor in the direction vector.
\item The flux linkage $\psi_f$ is also included in the same scalar factor and is therefore canceled when the rotor position is obtained from the direction of $\bm Y$.
\end{itemize}

As shown in Algorithm~\ref{alg:temporal_projection}, the choice of the resistance $\hat R_s$ is optional; when it is set to a nonzero value, it should not be chosen too large, so that $\kappa_R>0$ is preserved in \eqref{eq:kappa}.

\begin{figure}[!t]
    \centerline{\includegraphics[width=18em]{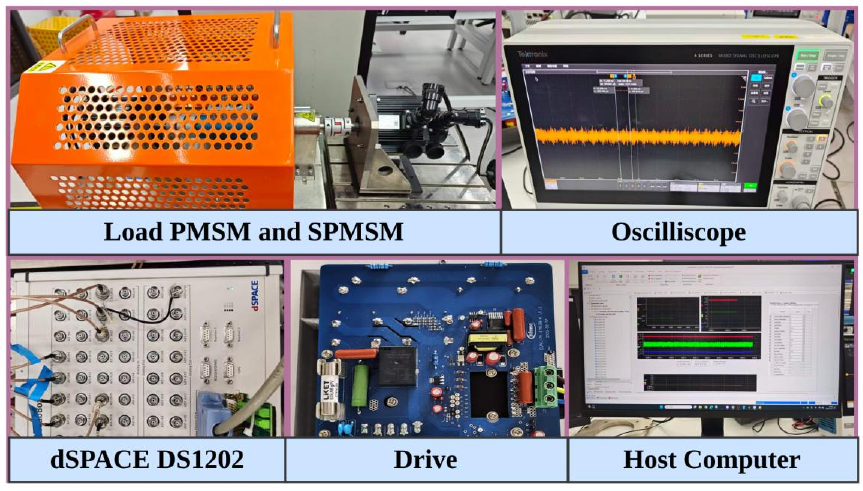}}
    \caption{Experimental verification platform.}
    \label{platform}
\end{figure} 

\begin{table}[!t]
  \captionsetup{labelsep=newline,justification=centering,singlelinecheck=false,font=footnotesize,labelfont=normalfont,textfont=sc,name=TABLE}
  \caption{Parameters of the SPMSM System}
  \renewcommand\arraystretch{1.25} 
  \small
  \centering
  \setlength{\tabcolsep}{5mm} 
  \begin{tabular}{c c c } \hline \hline
    Description & Parameter & Value   \\ \hline 
    Rated power & $P_r$ & 1 kW \\
    Rated torque & $T_r$ & 4 N $\!\cdot\!$ m \\
    Number of pole pairs & $N_p$ & 4  \\
    DC voltage & $U_{dc}$ & 300 V \\
    Sampling time & $T_s$ & 100 $\mu$s \\
    \hline
     Stator resistance & $R_s$ & 1.38 $\Omega$ \\
     Stator inductance & $L_s$ & 3.21 mH  \\
    Rotor flux linkage & $\psi_s $ & 0.148 Wb   \\ \hline
    Window size & $N$ & 30  \\
    Interval duration & $\Delta t$ & 2$T_s$ \\
    Algorithmic resistance & $\hat{R}_s$ & 0 \\
    \hline  \hline 
  \end{tabular}
  \label{Table.Parameters}
\end{table}

\begin{figure*}[!t]
\centering
\begin{subfigure}[t]{0.32\textwidth}
\centering
\includegraphics[width=\linewidth]{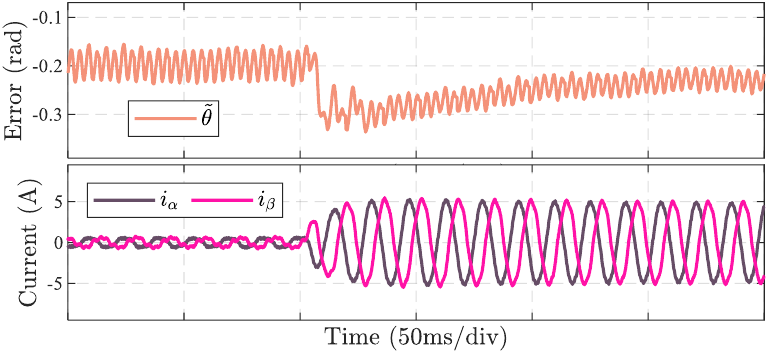}
\caption{Method 1}
\label{fig:free_1000}
\end{subfigure}
\hfill
\begin{subfigure}[t]{0.32\textwidth}
\centering
\includegraphics[width=\linewidth]{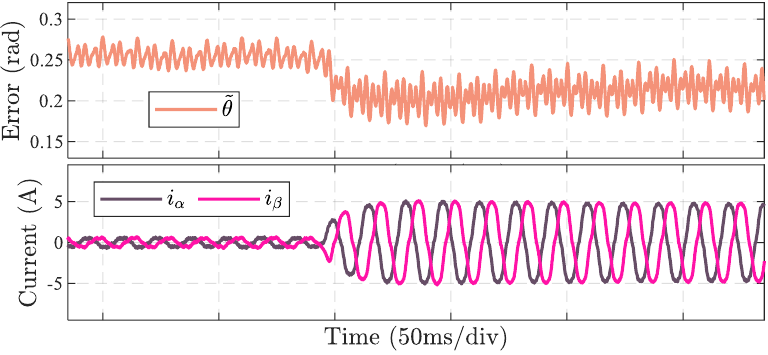}
\caption{Method 2}
\label{fig:smo_1000}
\end{subfigure}
\hfill
\begin{subfigure}[t]{0.32\textwidth}
\centering
\includegraphics[width=\linewidth]{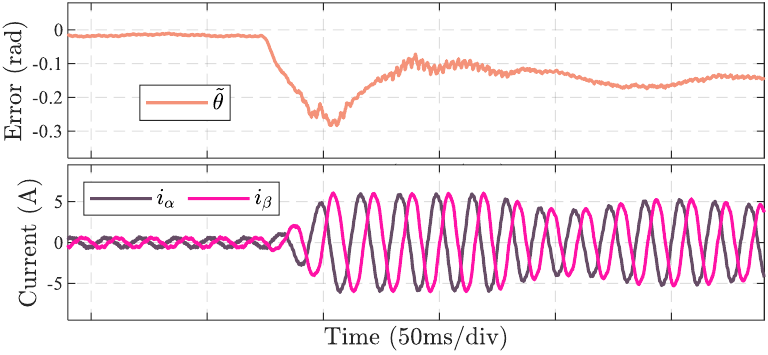}
\caption{Proposed TFC}
\label{fig:proposed_1000}
\end{subfigure}
\caption{Experimental results of different methods at 1000 rpm under a sudden full load transient.}
\label{fig:exp_1000}
\end{figure*}

\begin{figure*}[!t]
\centering
\begin{subfigure}[t]{0.24\textwidth}
\centering
\includegraphics[width=\linewidth]{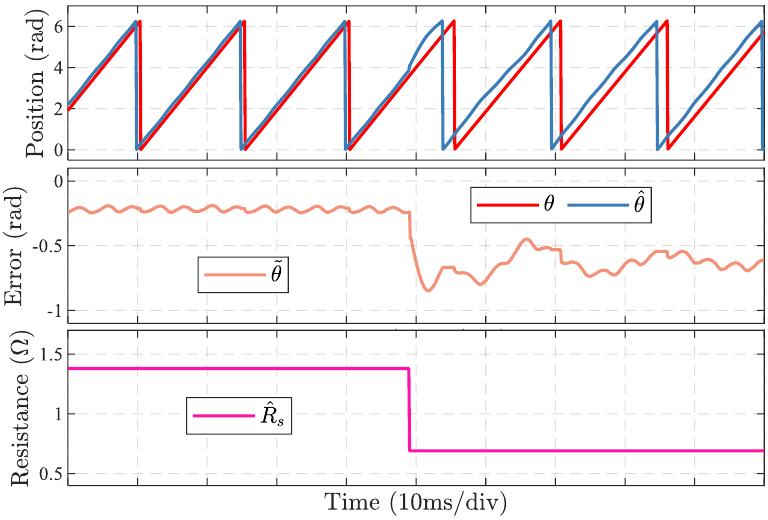}
\caption{$R_s \rightarrow 0.5R_s$}
\label{fig:free_1000_05r}
\end{subfigure}
\hfill
\begin{subfigure}[t]{0.24\textwidth}
\centering
\includegraphics[width=\linewidth]{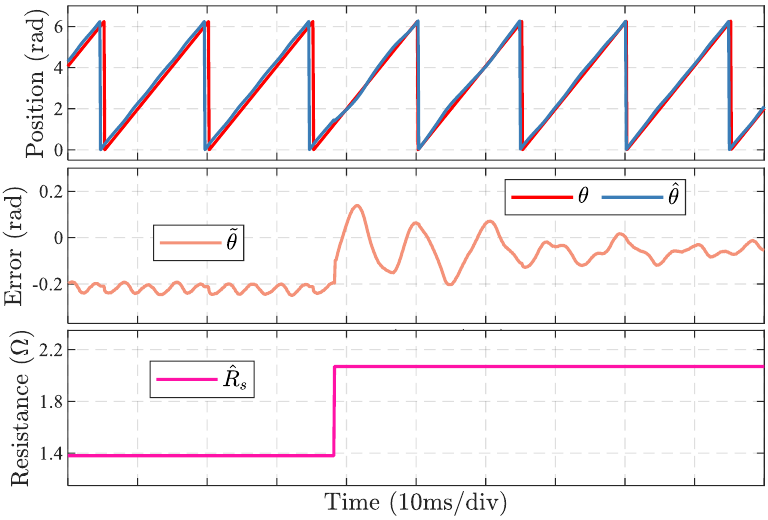}
\caption{$R_s \rightarrow 1.5R_s$}
\label{fig:free_1000_15r}
\end{subfigure}
\hfill
\begin{subfigure}[t]{0.24\textwidth}
\centering
\includegraphics[width=\linewidth]{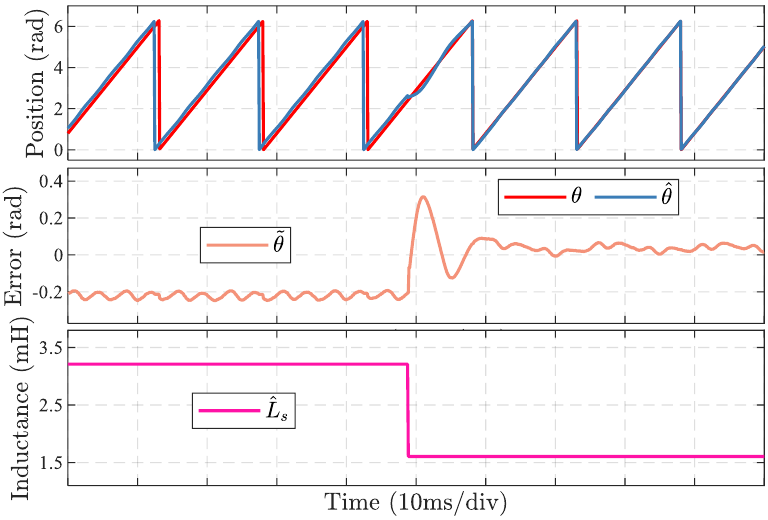}
\caption{$L_s \rightarrow 0.5L_s$}
\label{fig:free_1000_05l}
\end{subfigure}
\hfill
\begin{subfigure}[t]{0.24\textwidth}
\centering
\includegraphics[width=\linewidth]{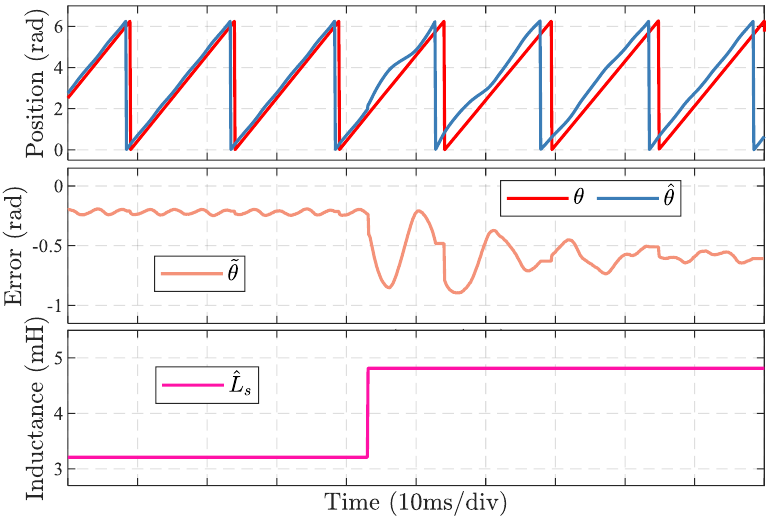}
\caption{$L_s \rightarrow 1.5L_s$}
\label{fig:free_1000_15l}
\end{subfigure}
\caption{Experimental results of the Method 1 under resistance and inductance variations at 1000 rpm and full load.}
\label{fig:free_1000_resistance}
\end{figure*}

\begin{figure*}[!t]
\centering
\begin{subfigure}[t]{0.24\textwidth}
\centering
\includegraphics[width=\linewidth]{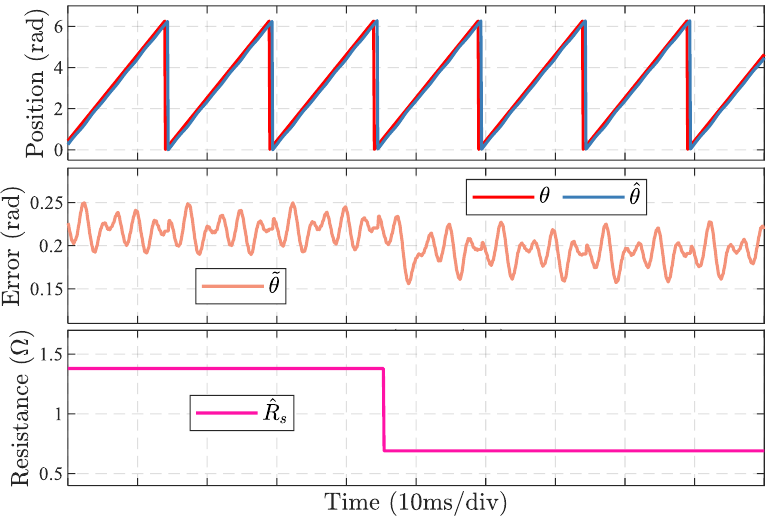}
\caption{$R_s \rightarrow 0.5R_s$}
\label{fig:SMO_1000_05r}
\end{subfigure}
\hfill
\begin{subfigure}[t]{0.24\textwidth}
\centering
\includegraphics[width=\linewidth]{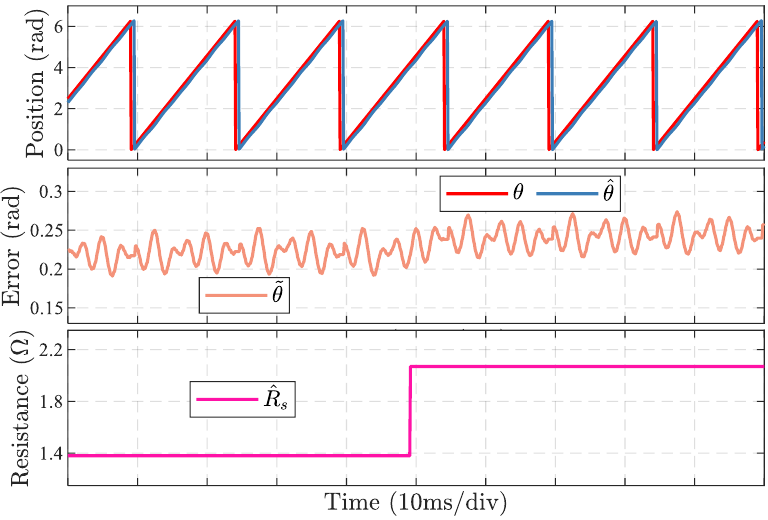}
\caption{$R_s \rightarrow 1.5R_s$}
\label{fig:SMO_1000_15r}
\end{subfigure}
\hfill
\begin{subfigure}[t]{0.24\textwidth}
\centering
\includegraphics[width=\linewidth]{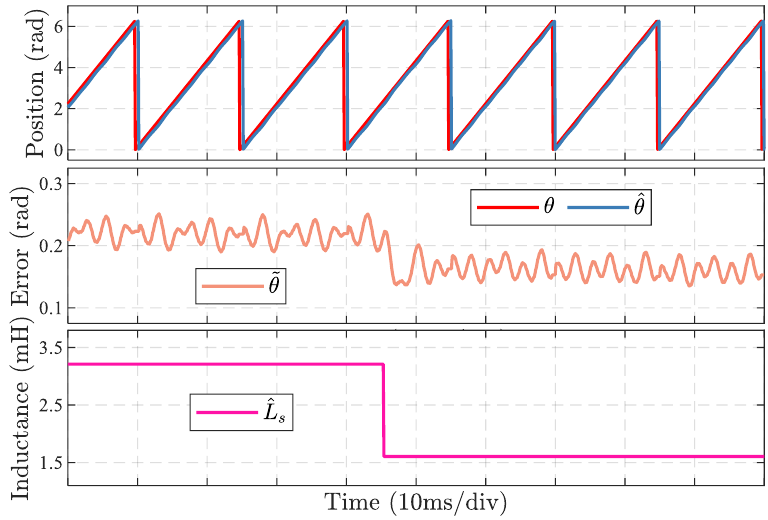}
\caption{$L_s \rightarrow 0.5L_s$}
\label{fig:SMO_1000_05l}
\end{subfigure}
\hfill
\begin{subfigure}[t]{0.24\textwidth}
\centering
\includegraphics[width=\linewidth]{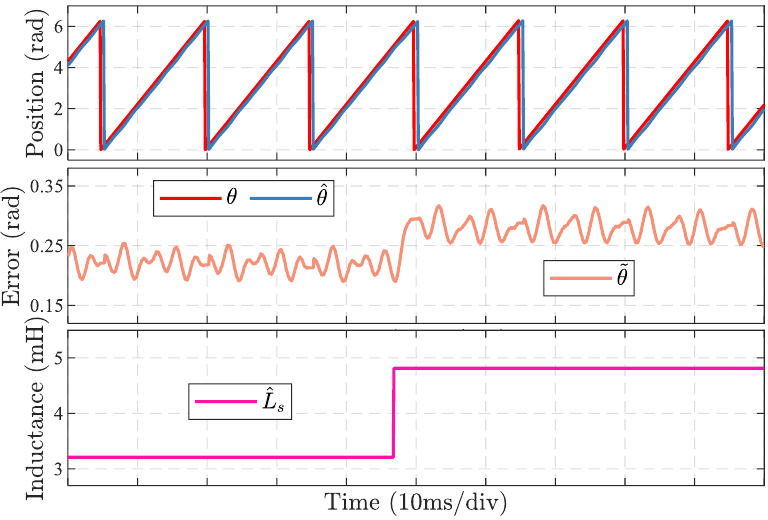}
\caption{$L_s \rightarrow 1.5L_s$}
\label{fig:SMO_1000_15l}
\end{subfigure}
\caption{Experimental results of the Method 2 under resistance and inductance variations at 1000 rpm and full load.}
\label{fig:SMO_1000_resistance}
\end{figure*}

\section{Experimental Results and Analysis}
\label{sec:exp_results}

\subsection{Experimental Setup}

The proposed TFC method is evaluated on the SPMSM platform shown in Fig.~\ref{platform}. Two mechanically coupled PMSMs form the drive-load arrangement, with control implemented on a dSPACE DS1202 system connected to a power driver and a host computer. The speed controller is updated at 2 kHz, while the current controller operates at 10 kHz. Table~\ref{Table.Parameters} summarizes the motor and control parameters, including the window size $N$ and interval duration $\Delta t$ used by the proposed TFC method. Two approaches are selected as benchmarks: Method 1 is a model-free controller in \cite{11204721}, and Method 2 is the SMO-based method in \cite{10557457}.

\makeatletter
\setlength{\@dblfptop}{0pt}
\makeatother

\subsection{Performance under Load Transition}

Fig. \ref{fig:exp_1000} compares the load-transition performance of the three methods at 1000 rpm. Before loading, the proposed TFC maintains a position error of about \(-0.02\) rad, which is smaller than those of Method 1 and Method 2, at approximately \(-0.19\) rad and \(0.25\) rad, respectively. After the rated load is suddenly applied, all three methods remain stable. The proposed method settles to about \(-0.15\) rad, whereas Method 1 and Method 2 exhibit steady-state errors of approximately \(-0.23\) rad and \(0.21\) rad, respectively. In addition, Method 1 and Method 2 exhibit more pronounced error oscillations, whereas the proposed TFC method provides smoother position estimation.

\begin{figure*}[!t]
\centering
\begin{subfigure}[t]{0.24\textwidth}
\centering
\includegraphics[width=\linewidth]{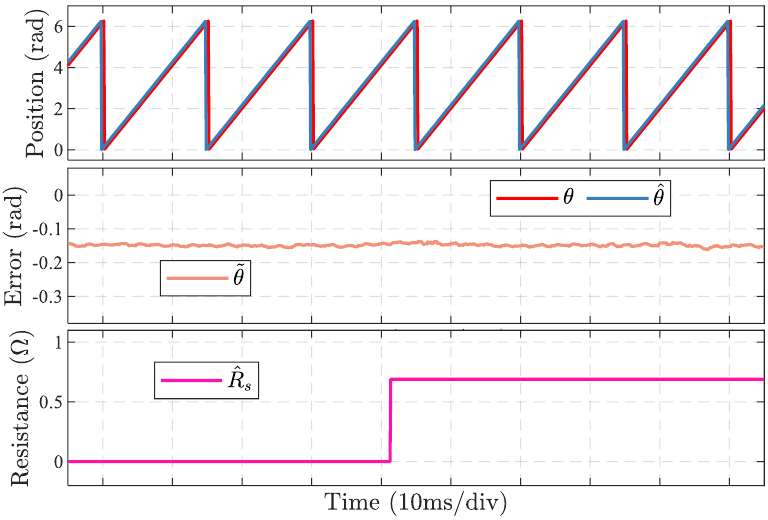}
\caption{$0 \rightarrow 0.5R_s$}
\label{fig:proposed_1000_05r}
\end{subfigure}
\hfill
\begin{subfigure}[t]{0.24\textwidth}
\centering
\includegraphics[width=\linewidth]{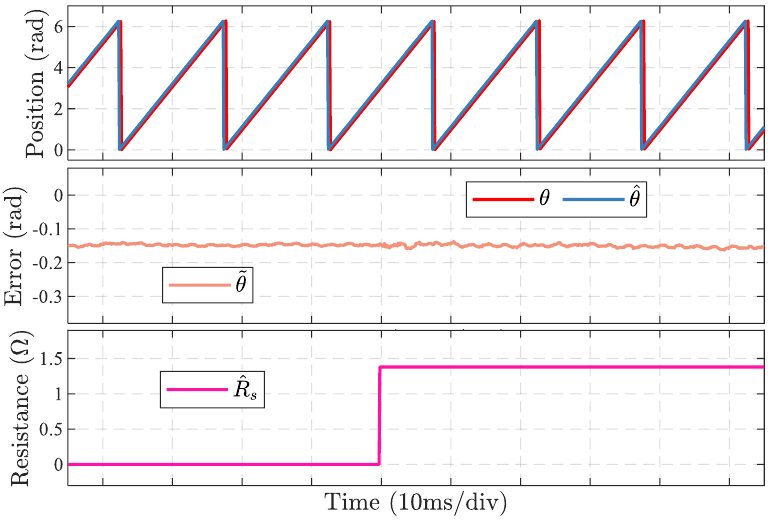}
\caption{$0 \rightarrow R_s$}
\label{fig:proposed_1000_r}
\end{subfigure}
\hfill
\begin{subfigure}[t]{0.24\textwidth}
\centering
\includegraphics[width=\linewidth]{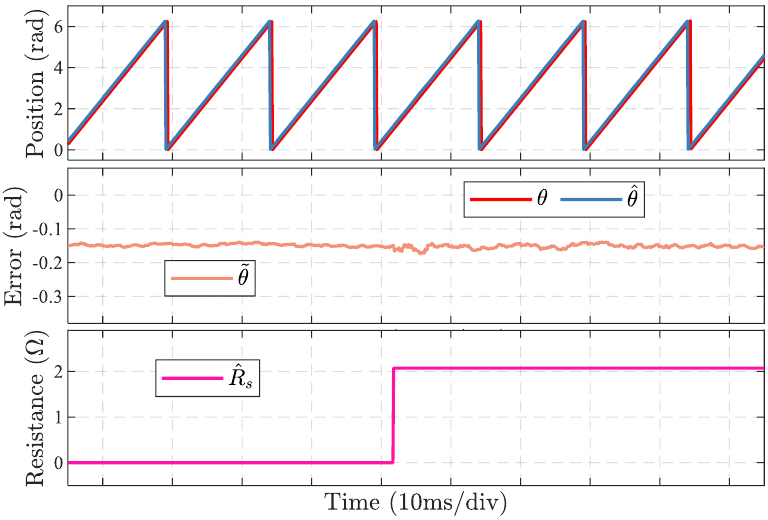}
\caption{$0 \rightarrow 1.5R_s$}
\label{fig:proposed_1000_15r}
\end{subfigure}
\hfill
\begin{subfigure}[t]{0.24\textwidth}
\centering
\includegraphics[width=\linewidth]{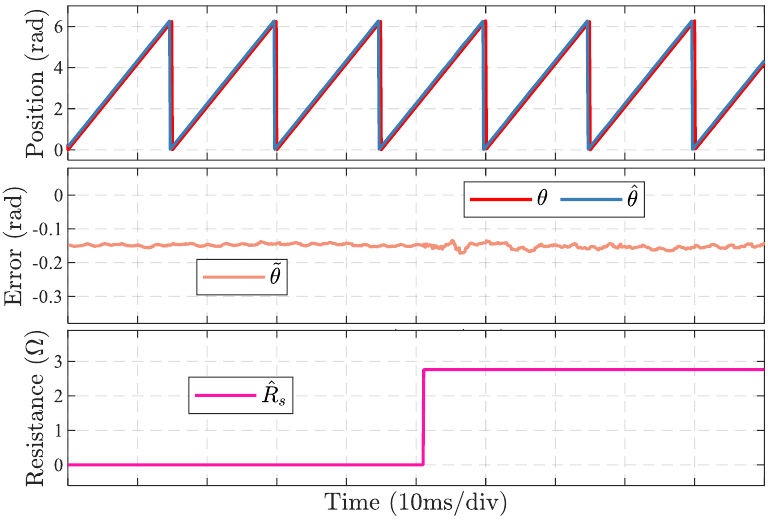}
\caption{$0 \rightarrow 2R_s$}
\label{fig:proposed_1000_2r}
\end{subfigure}
\caption{Experimental results of the proposed TFC method under resistance variations at 1000 rpm and full load.}
\label{fig:proposed_1000_resistance}
\end{figure*}

\subsection{Performance under Parameter Variations}
\begin{table}[!t]
\centering
\captionsetup{labelsep=newline,justification=centering,singlelinecheck=false,font=footnotesize,labelfont=normalfont,textfont=sc,name=TABLE}
\caption{RMSE of Estimation Errors under Different Parameter Mismatch Conditions at 1000 rpm. Unit: rad}
\label{tab:parameter_errors_1000}
\small
\setlength{\tabcolsep}{0pt}
\renewcommand{\arraystretch}{1.25}
\begin{tabular}{>{\centering\arraybackslash}p{0.265\columnwidth}*{4}{>{\centering\arraybackslash}p{0.17\columnwidth}}}
\hline\hline
 & 0.5$R_s$ & 1.5$R_s$ & 0.5$L_s$ & 1.5$L_s$ \\
\hline
Method 1 & 0.65 & 0.05 & 0.03 & 0.60 \\
Method 2 & 0.19 & 0.24 & 0.16 & 0.28 \\
\hline
 & 0.5$R_s$ & $R_s$ & 1.5$R_s$ & 2$R_s$ \\
\hline
Proposed TFC & 0.15 & 0.15 & 0.15 & 0.15 \\
\hline\hline
\end{tabular}
\begin{minipage}{\columnwidth}
\end{minipage}
\end{table}

To evaluate the parameter sensitivity, step variations of the resistance and inductance are introduced at 1000 rpm under full load. As shown in Fig. \ref{fig:free_1000_resistance}, Method 1 exhibits significant sensitivity to both parameters. Under \(0.5R_s\), the position error changes from about \(-0.22\) rad to \(-0.65\) rad, while \(1.5L_s\) results in an error of about \(-0.60\) rad.   As shown in Fig. \ref{fig:SMO_1000_resistance}, Method 2 is less sensitive to parameter variations, with only small shifts in the position error under resistance and inductance changes.   As shown in Fig. \ref{fig:proposed_1000_resistance}, although the proposed TFC method does not require motor parameters, the resistance is optional and can be selected.  Therefore, \(\hat{R}_s\) is varied from zero to \(0.5R_s\), \(R_s\), \(1.5R_s\), and \(2R_s\). The position error remains around \(-0.15\) rad in all cases, with only slight transient fluctuations. Overall, the proposed TFC method exhibits the strongest robustness against parameter variations. Table~\ref{tab:parameter_errors_1000} summarizes the root mean square error (RMSE) of the position estimate for the three methods under parameter mismatch.

\begin{figure*}[!t]
\centering
\begin{subfigure}[t]{0.24\textwidth}
\centering
\includegraphics[width=\linewidth]{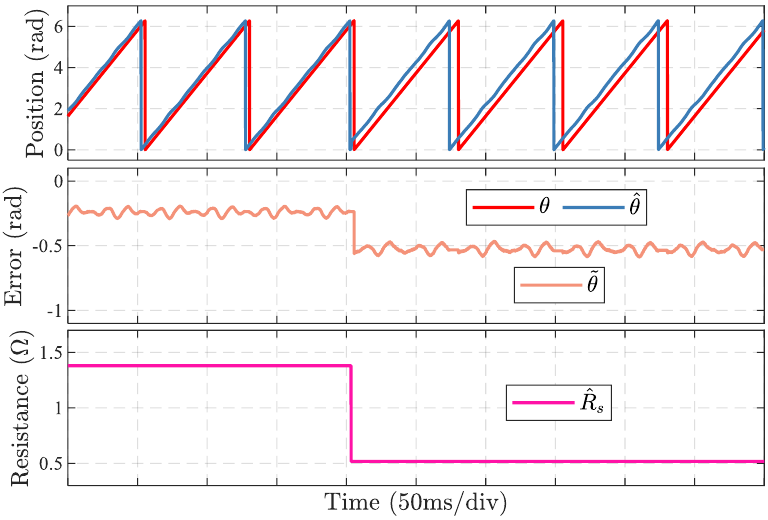}
\caption{$R_s \rightarrow 0.5R_s$}
\label{fig:free_200_05r}
\end{subfigure}
\hfill
\begin{subfigure}[t]{0.24\textwidth}
\centering
\includegraphics[width=\linewidth]{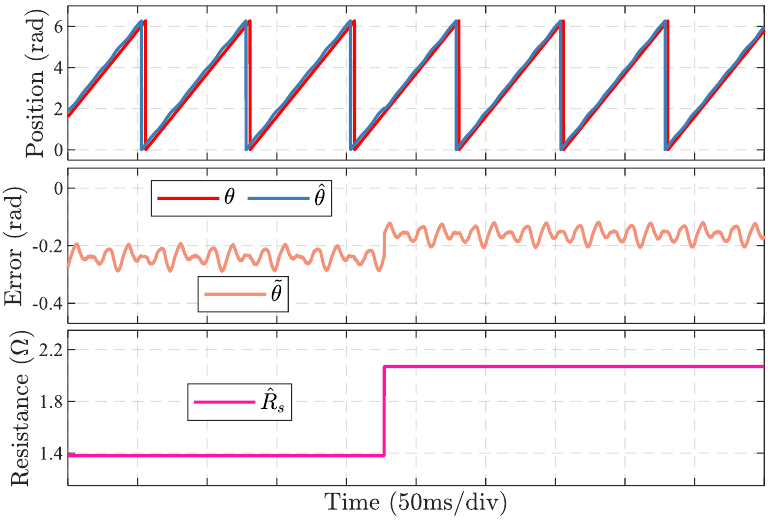}
\caption{$R_s \rightarrow 1.5R_s$}
\label{fig:free_200_15r}
\end{subfigure}
\hfill
\begin{subfigure}[t]{0.24\textwidth}
\centering
\includegraphics[width=\linewidth]{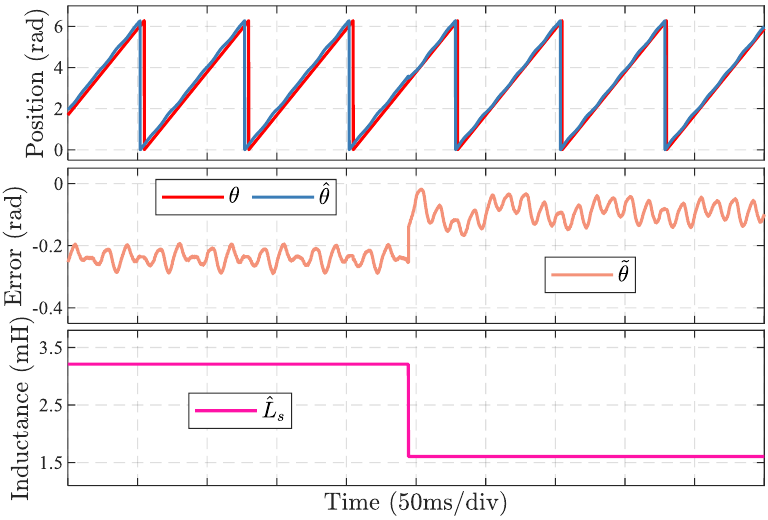}
\caption{$L_s \rightarrow 0.5L_s$}
\label{fig:free_200_05l}
\end{subfigure}
\hfill
\begin{subfigure}[t]{0.24\textwidth}
\centering
\includegraphics[width=\linewidth]{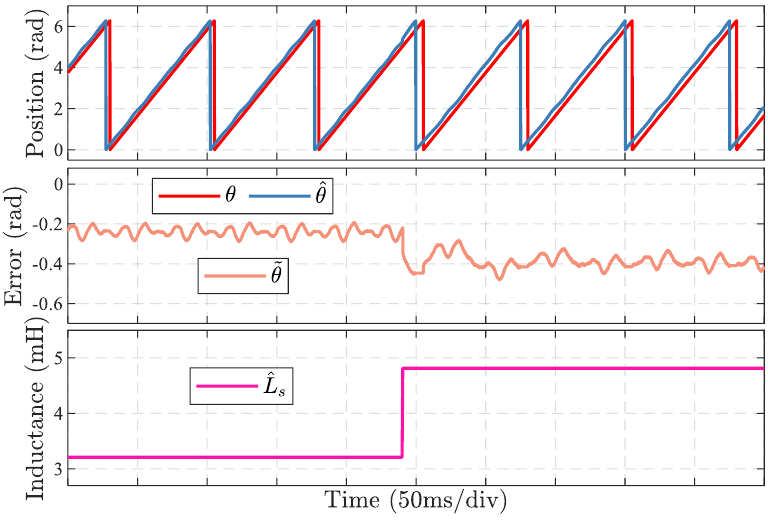}
\caption{$L_s \rightarrow 1.5L_s$}
\label{fig:free_200_15l}
\end{subfigure}
\caption{Experimental results of the Method 1 under resistance and inductance variations at 200 rpm and full load.}
\label{fig:free_200_resistance}
\end{figure*}

\begin{figure*}[!t]
\centering
\begin{subfigure}[t]{0.24\textwidth}
\centering
\includegraphics[width=\linewidth]{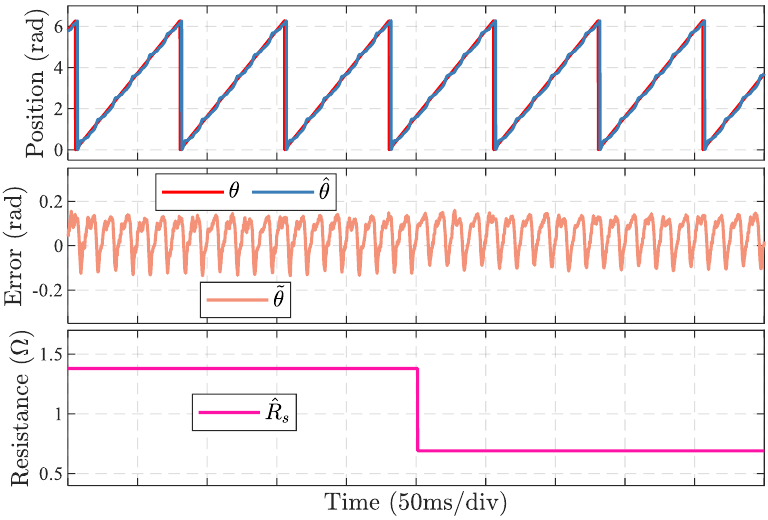}
\caption{$R_s \rightarrow 0.5R_s$}
\label{fig:SMO_200_05r}
\end{subfigure}
\hfill
\begin{subfigure}[t]{0.24\textwidth}
\centering
\includegraphics[width=\linewidth]{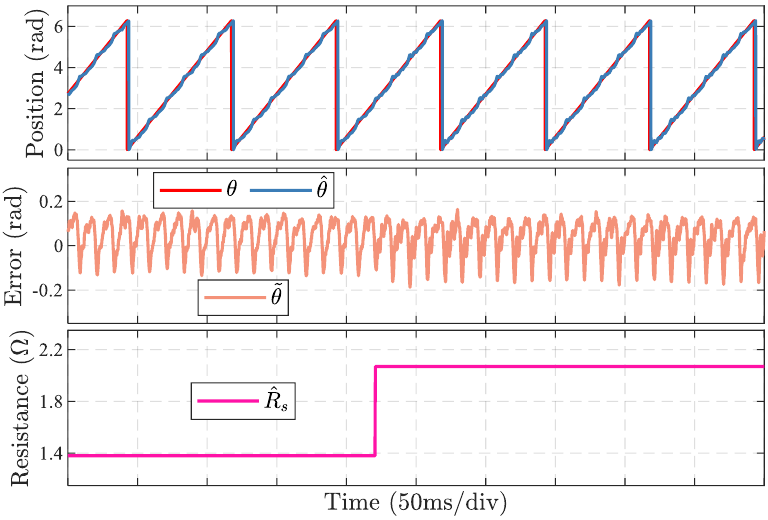}
\caption{$R_s \rightarrow 1.5R_s$}
\label{fig:SMO_200_15r}
\end{subfigure}
\hfill
\begin{subfigure}[t]{0.24\textwidth}
\centering
\includegraphics[width=\linewidth]{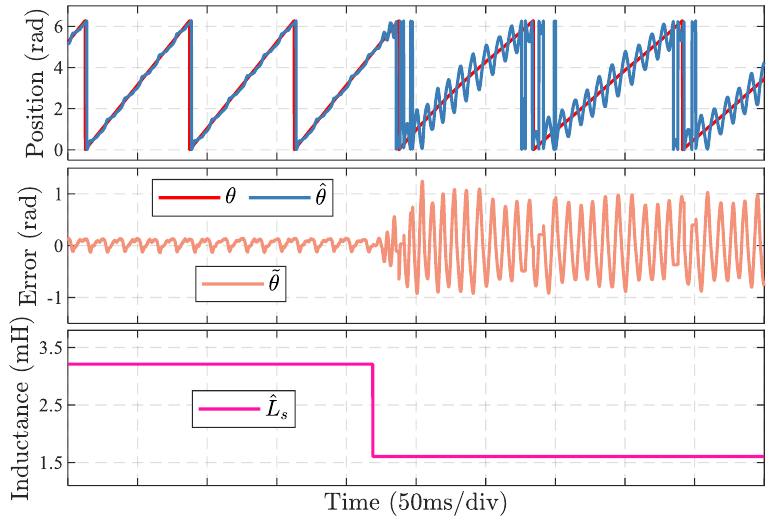}
\caption{$L_s \rightarrow 0.5L_s$}
\label{fig:SMO_200_05l}
\end{subfigure}
\hfill
\begin{subfigure}[t]{0.24\textwidth}
\centering
\includegraphics[width=\linewidth]{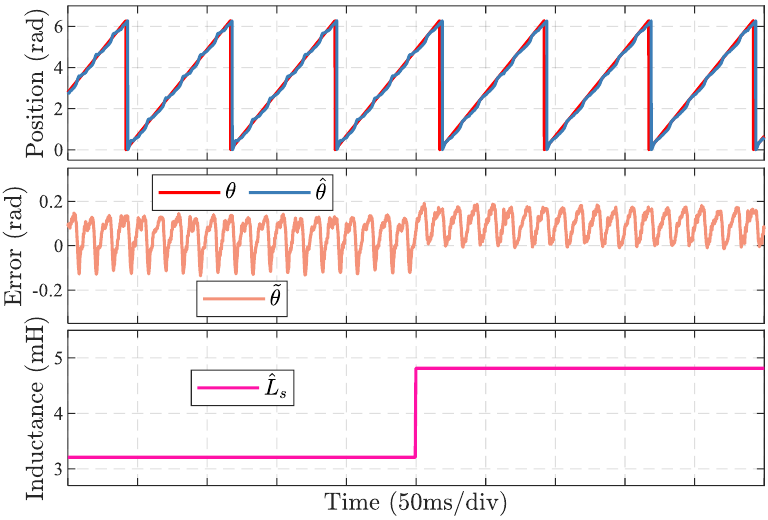}
\caption{$L_s \rightarrow 1.5L_s$}
\label{fig:SMO_200_15l}
\end{subfigure}
\caption{Experimental results of the Method 2 under resistance and inductance variations at 200 rpm and full load.}
\label{fig:SMO_200_resistance}
\end{figure*}

\subsection{Performance under Low-Speed Operation}

To evaluate low-speed parameter robustness, experiments are conducted at 200 rpm under full load, with the stator resistance and inductance stepped to \(0.5R_s\), \(1.5R_s\), \(0.5L_s\), and \(1.5L_s\), as shown in Figs.~\ref{fig:free_200_resistance} and \ref{fig:SMO_200_resistance}. Method 1 exhibits smaller error shifts than at 1000 rpm, with errors of approximately $-0.52$ rad at $0.5R_s$ and $-0.40$ rad at $1.5L_s$. Method 2 is particularly sensitive to inductance reduction, with error oscillations reaching approximately $\pm 1.0$ rad at $0.5L_s$.

\begin{figure*}[t]
\centering
\begin{subfigure}[t]{0.24\textwidth}
\centering
\includegraphics[width=\linewidth]{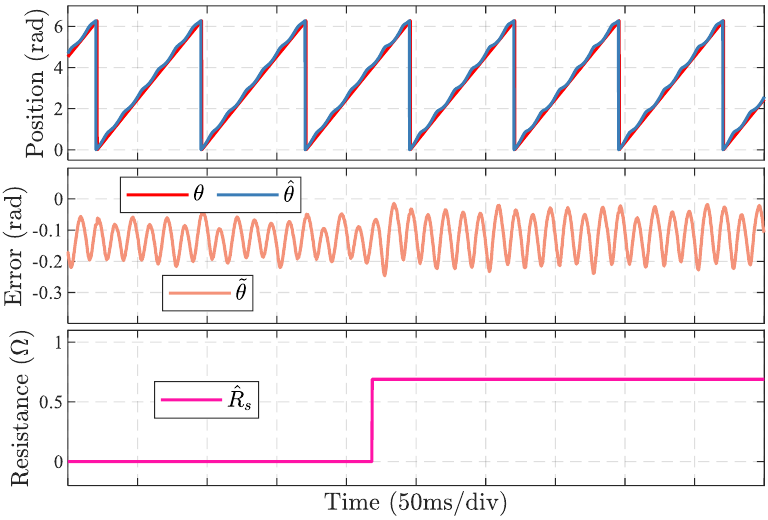}
\caption{$0 \rightarrow 0.5R_s$}
\label{fig:proposed_200_05r}
\end{subfigure}
\hfill
\begin{subfigure}[t]{0.24\textwidth}
\centering
\includegraphics[width=\linewidth]{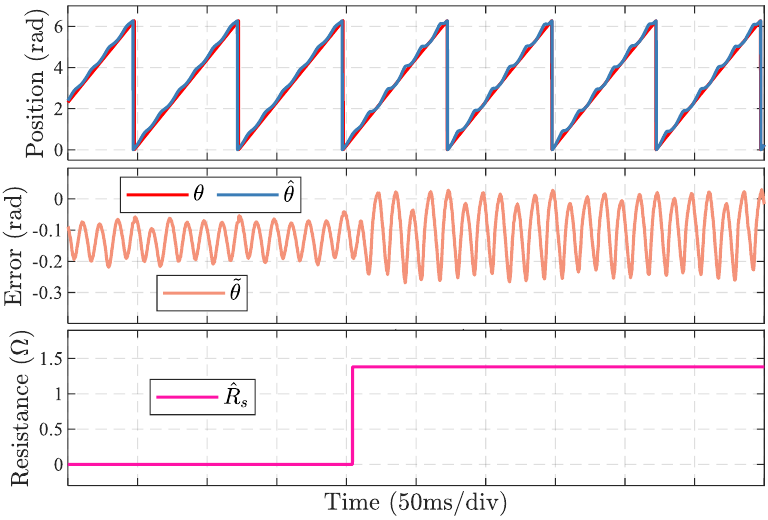}
\caption{$0 \rightarrow R_s$}
\label{fig:proposed_200_r}
\end{subfigure}
\hfill
\begin{subfigure}[t]{0.24\textwidth}
\centering
\includegraphics[width=\linewidth]{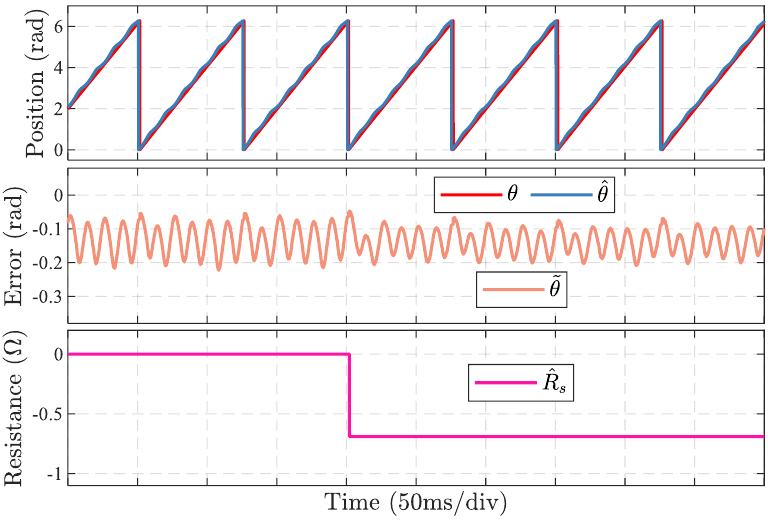}
\caption{$0 \rightarrow -0.5R_s$}
\label{fig:proposed_200_-0.5r}
\end{subfigure}
\hfill
\begin{subfigure}[t]{0.24\textwidth}
\centering
\includegraphics[width=\linewidth]{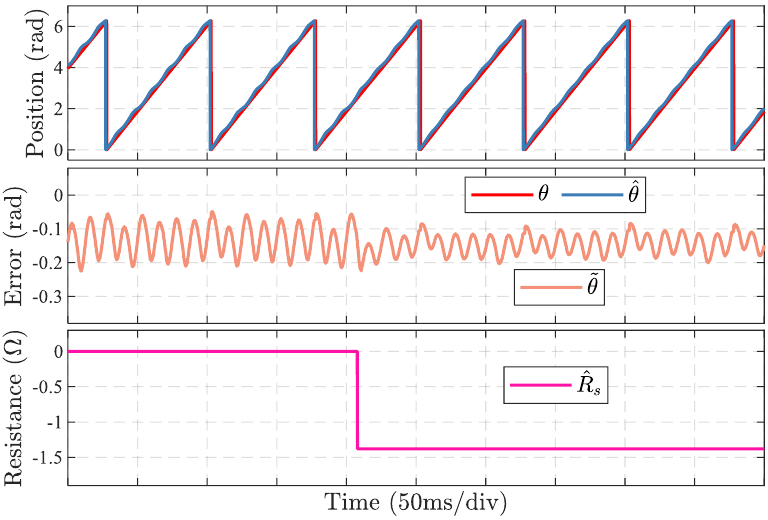}
\caption{$0 \rightarrow -R_s$}
\label{fig:proposed_200_-R}
\end{subfigure}
\caption{Experimental results of the proposed TFC method under resistance variations at 200 rpm and full load.}
\label{fig:proposed_200_resistance}
\end{figure*}

Different parameter variations are applied to the proposed TFC method in Fig.~\ref{fig:proposed_200_resistance}, where the resistance \(\hat R_s\) is tested using both positive and negative values. According to \eqref{eq:kappa}, \(\kappa_R\) should remain positive, and a negative \(\hat R_s\) can increase \(\kappa_R\) and improve low-speed observability. Variations in \(\hat R_s\) have little effect on the mean error but change the error oscillation.

\begin{figure*}[!t]
\centering
\begin{subfigure}[t]{0.32\textwidth}
\centering
\includegraphics[width=\linewidth]{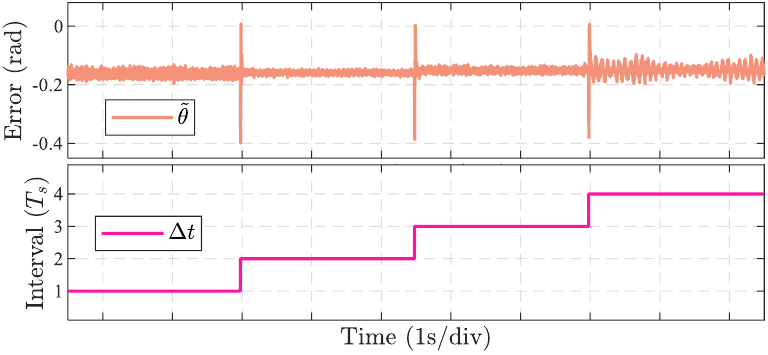}
\caption{$N=20$}
\label{fig:Proposed_1000_N_20}
\end{subfigure}
\hfill
\begin{subfigure}[t]{0.32\textwidth}
\centering
\includegraphics[width=\linewidth]{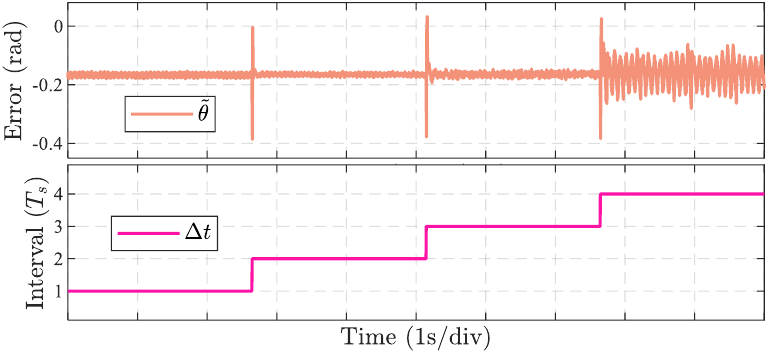}
\caption{$N=30$}
\label{fig:Proposed_1000_N_30}
\end{subfigure}
\hfill
\begin{subfigure}[t]{0.32\textwidth}
\centering
\includegraphics[width=\linewidth]{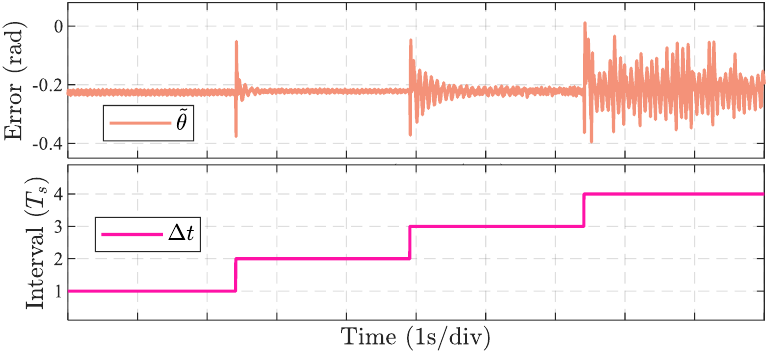}
\caption{$N=40$}
\label{fig:Proposed_1000_N_40}
\end{subfigure}
\caption{Experimental result of the proposed TFC method under different window sizes and interval durations at 1000 rpm and full load.}
\label{fig:exp_1000_N}
\end{figure*}

\begin{figure*}[!t]
\centering
\begin{subfigure}[t]{0.32\textwidth}
\centering
\includegraphics[width=\linewidth]{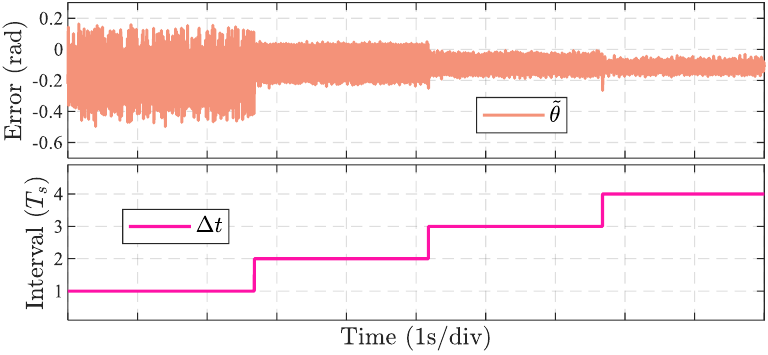}
\caption{$N=20$}
\label{fig:Proposed_200_N_20}
\end{subfigure}
\hfill
\begin{subfigure}[t]{0.32\textwidth}
\centering
\includegraphics[width=\linewidth]{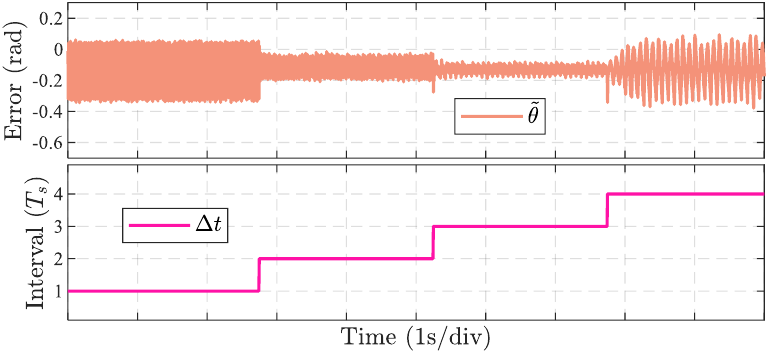}
\caption{$N=30$}
\label{fig:Proposed_200_N_30}
\end{subfigure}
\hfill
\begin{subfigure}[t]{0.32\textwidth}
\centering
\includegraphics[width=\linewidth]{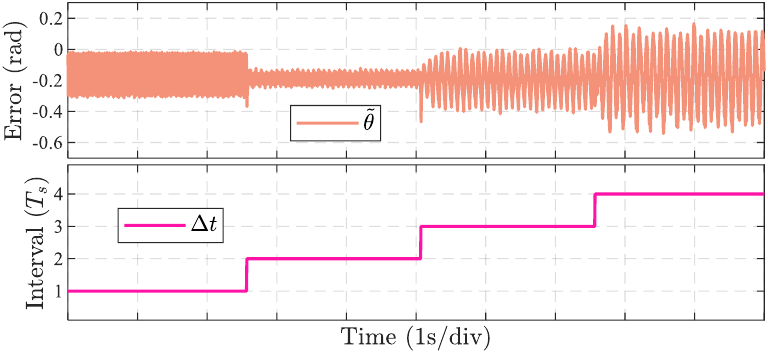}
\caption{$N=40$}
\label{fig:Proposed_200_N_40}
\end{subfigure}
\caption{Experimental result of the proposed TFC method under different window sizes and interval durations at 200 rpm and full load.}
\label{fig:exp_200_N}
\end{figure*}

\subsection{Performance under Regression Parameter Variations}
To investigate the influence of the regression parameters, the window size \(N\) and interval duration \(\Delta t\) are varied at 1000 and 200 rpm. As shown in Fig. \ref{fig:exp_1000_N}, at 1000 rpm, a small or moderate \(\Delta t\) provides stable estimation, whereas the error oscillation increases as \(\Delta t\) and \(N\) become large, especially for \(N=40\). At 200 rpm, Fig. \ref{fig:exp_200_N} shows that an excessively small \(\Delta t\) may lead to insufficient rotational excitation, while increasing \(\Delta t\) initially reduces the estimation oscillation. However, an overly large interval or window again increases the error oscillation. In practice, a smaller interval duration can be selected at medium and high speeds, whereas a larger interval duration is preferred at low speeds to preserve sufficient rotational excitation.

\begin{figure}[!t]
    \centerline{\includegraphics[width=24em]{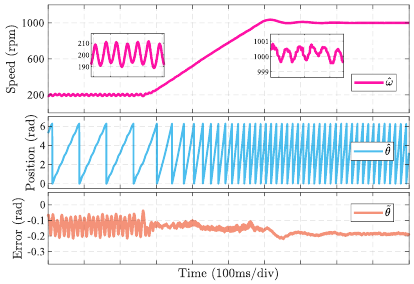}}
    \caption{Experimental results during acceleration from 200 rpm to 1000 rpm.}
    \label{acceleration}
\end{figure}

\subsection{Performance under Speed Transients}
To evaluate the dynamic performance over a wide speed range, the motor accelerates from 200 to 1000 rpm. As shown in Fig. \ref{acceleration}, the proposed TFC method maintains continuous position estimation throughout the acceleration process without loss of synchronization. The speed fluctuation is about 10 rpm at 200 rpm and decreases to approximately 1 rpm at 1000 rpm, while the position error remains bounded during the entire transition.

Overall, the experimental results show that the proposed TFC method achieves model-free sensorless control with few tuning parameters, strong robustness, and reliable operation over a wide-speed range through purely data-driven estimation.

\section{Conclusion}
This paper proposes a temporal regression-based model-free sensorless control method for SPMSM. By constructing a finite window regression from measured voltage and current data, the inductance term is eliminated, while the effects of resistance and flux linkage are canceled during position extraction. Thus, position estimation is achieved without requiring resistance, inductance, or flux linkage. Experimental results demonstrate strong robustness against load and parameter variations, reliable low-speed operation, and stable performance from 200 to 1000 rpm. The effects of window size and interval duration are also investigated to provide practical tuning guidelines. In summary, the proposed TFC offers a purely data-driven sensorless solution with few tuning parameters and reliable wide-speed range operation.

\bibliographystyle{IEEEtran}
\bibliography{references}


\vspace{-0.3cm} 
\begin{IEEEbiography}[{\includegraphics[width=1in,height=1.25in,clip,keepaspectratio]{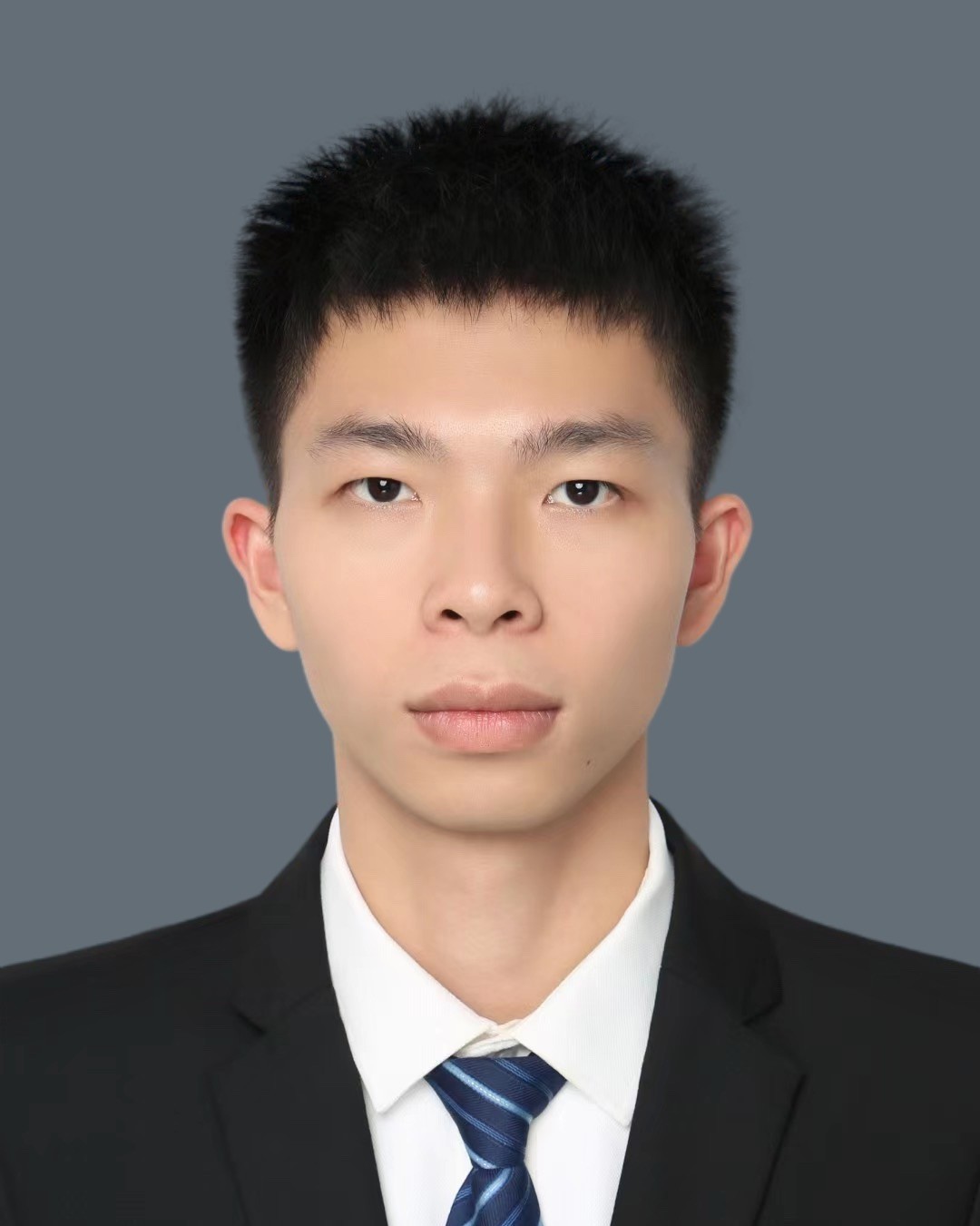}}]
{Fobao Zhou} (Graduate Student Member, IEEE) received the B.Eng. degree in Robot Engineering from Guangzhou University, Guangzhou, China, in 2022. He is currently pursuing the Ph.D. degree at The Hong Kong University of Science and Technology (Guangzhou), Guangzhou, China. His research interests include the integration of machine learning and control theory for the development of advanced intelligent systems. He aims to contribute to the field of engineering through innovative research and practical applications in these areas.
\end{IEEEbiography}

\vspace{-0.3cm}
\begin{IEEEbiography}[{\includegraphics[width=1in,height=1.25in,clip,keepaspectratio]{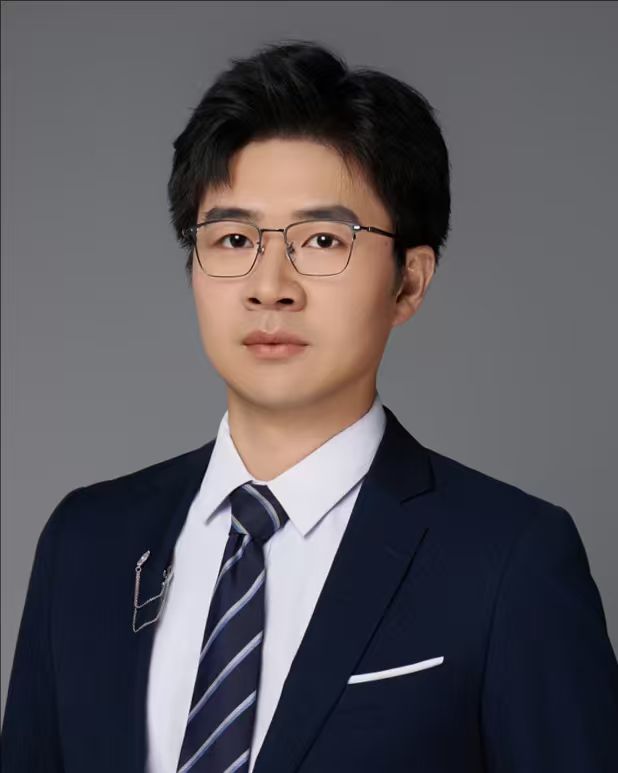}}]{Yang Shen} (Graduate Student Member, IEEE) received the B.S. and M.S. degrees from the North China University of Technology, Beijing, China, in 2018 and 2021, respectively. He is currently working toward the Ph.D. degree in Robotics and Autonomous Systems at the
Hong Kong University of Science and Technology (HKUST), Guangzhou and Hong Kong SAR, China. He is also a full-time cross-campus PhD student in the Department of Electronic and Computer Engineering in HKUST. His research interests include the research of special motor control and linear motor drive.
 \end{IEEEbiography}

\vspace{-0.3cm}
\begin{IEEEbiography}[{\includegraphics[width=1in,height=1.25in,clip,keepaspectratio]{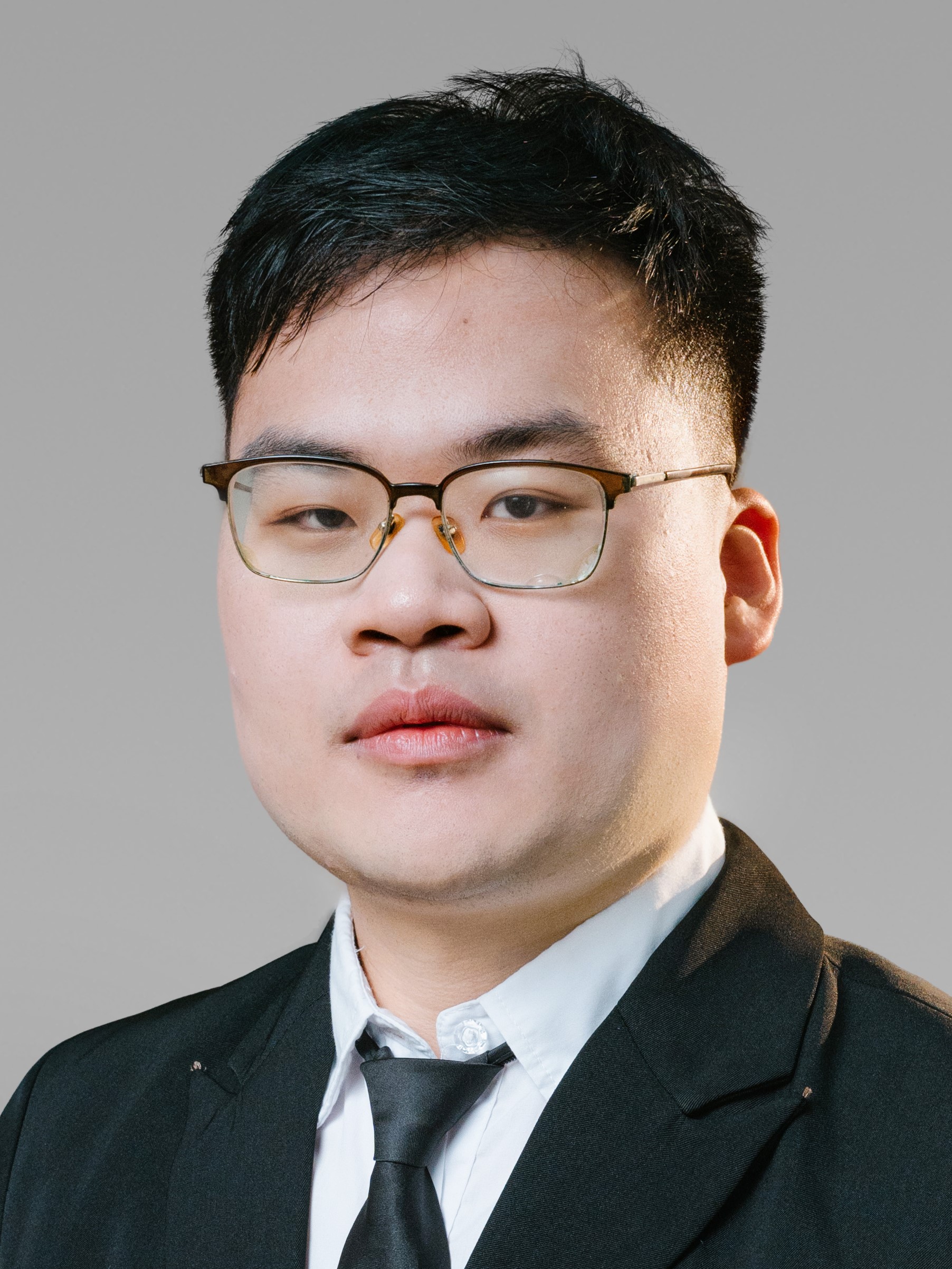}}]
{Xueyan Wang} received the B.S. degree in electrical engineering and automation and the M.S. degree in electrical engineering from Xi’an University of Science and Technology, Xi’an, China, in 2020 and 2023, respectively. He is currently pursuing the Ph.D. degree at The Hong Kong University of Science and Technology (Guangzhou), Guangzhou, China. His research interest covers model predictive control of permanent magnet synchronous motor drives, power electronics, and power transmission.
\end{IEEEbiography}

\vspace{-0.3cm}
\begin{IEEEbiography}
[{\includegraphics[width=1in,height=1.25in,clip,keepaspectratio]{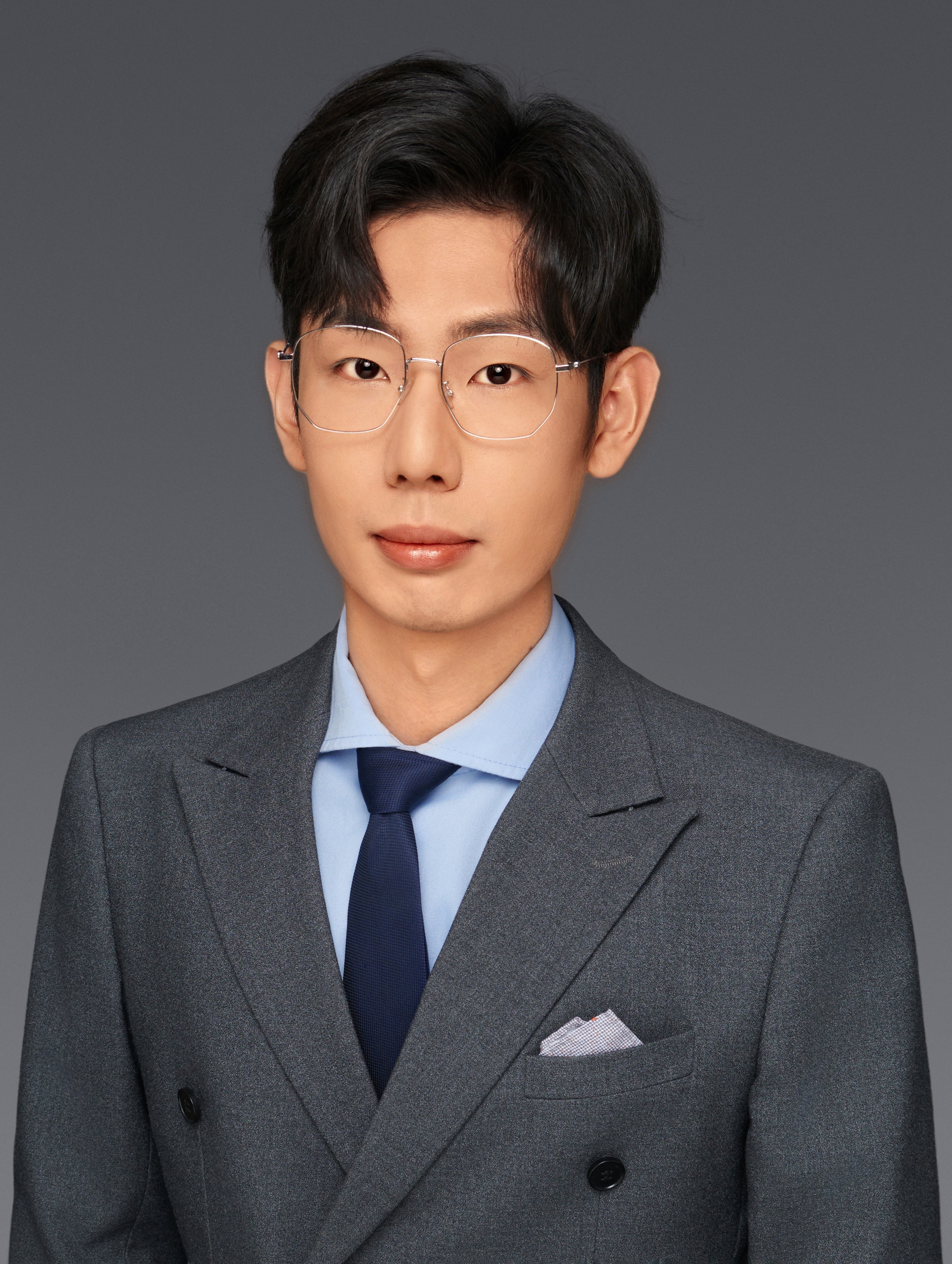}}]{Zhenxiao Yin} (Graduate Student Member, IEEE) received the B.Sc. degree in mechanical engineering from Paderborn University (UPB), Paderborn, Germany, in 2018, and the double M.Sc. degrees in mechanical engineering, mechatronics and robotics from the Technical University of Munich (TUM), Munich, Germany, in 2021 and 2022, respectively. He received the Ph.D. degree in Robotics and Autonomous Systems from The Hong Kong University of Science and Technology (HKUST), Guangzhou and Hong Kong SAR, China, in 2026. He was also a full-time cross-campus PhD student in the Department of Mechanical and Aerospace Engineering in HKUST. 

His research interests include learning-based control and state drop compensator for electric machines in flying vehicles.

He was the recipient of German National Scholarship from TUM. 
 \end{IEEEbiography}

\vspace{-0.3cm}
\begin{IEEEbiography}
[{\includegraphics[width=1in,height=1.25in,clip,keepaspectratio]{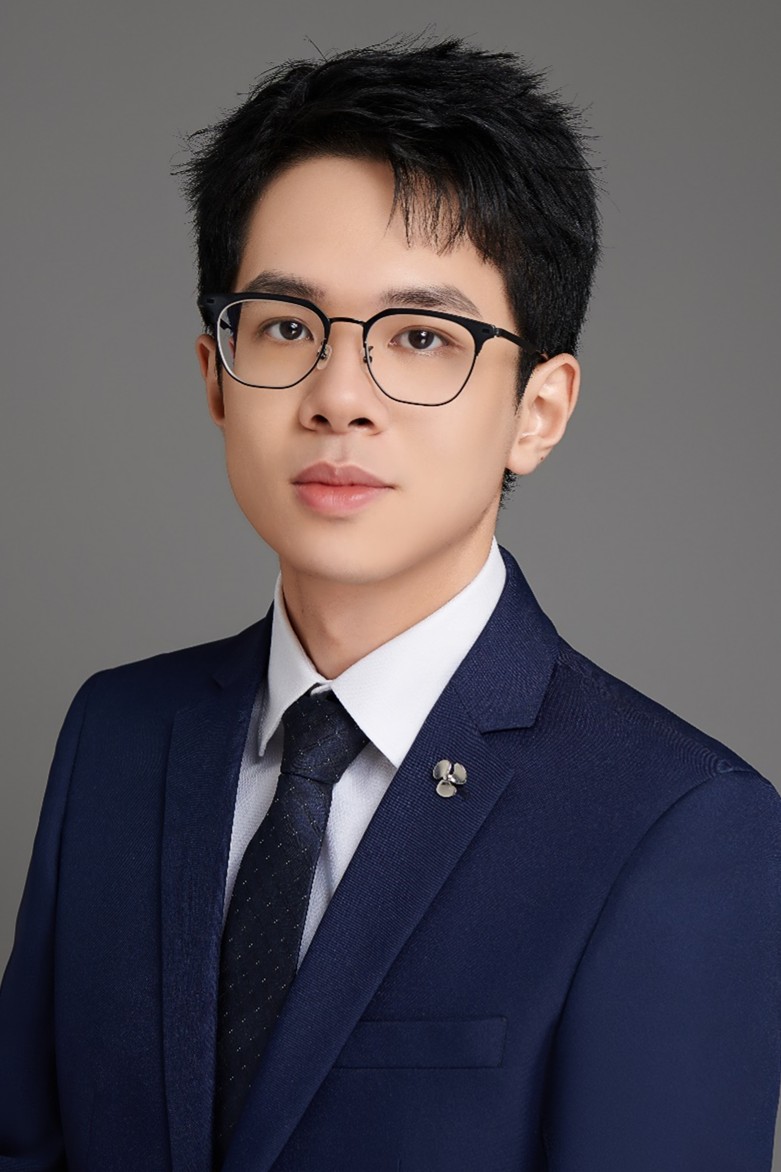}}]{Yujia ZHANG} received the B.Eng. degree in mechanical engineering from South China University of Technology, Guangzhou, China, in 2023, and received the M.Phil. in Robotics and Autonomous Systems at the Hong Kong University of Science and Technology (Guangzhou), Guangzhou. He is currently pursuing Ph.D. at the Hong Kong University of Science and Technology (Guangzhou).  His research interests include control of multi-phase machines, robotic joint control techniques, online neural network techniques and signal processing.
If you want more information, check https://orcid.org/0009-0000-9392-1041
 \end{IEEEbiography}


\begin{IEEEbiography}
[{\includegraphics[width=1in,height=1.25in,clip,keepaspectratio]{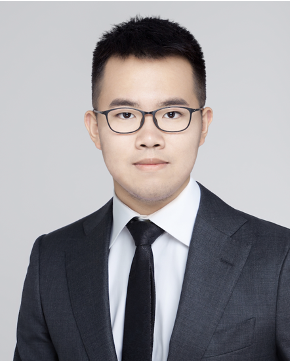}}]{Yuanfeng Qu} 
received the B.Sc. degree in Automation from Beijing Institute of Technology, Beijing, China, in 2023, where he was a recipient of the School Scholarship. He received the M.Sc. degree in Electronic Engineering from The Hong Kong University of Science and Technology, Hong Kong, China, in 2024. He is currently pursuing the Ph.D. degree in Robotics and Autonomous Systems at The Hong Kong University of Science and Technology (Guangzhou), Guangzhou, China. His research interests include learning-based control, advanced motor control, and autonomous systems.
 \end{IEEEbiography}

\vspace{-14cm}
\begin{IEEEbiography}[{\includegraphics[width=1in,height=1.25in,clip,keepaspectratio]{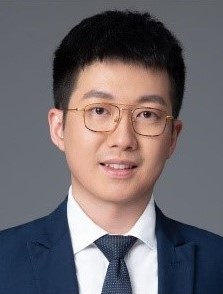}}]
{Hang Zhao} (Member, IEEE) received the B.Eng. and M.Eng. degrees from Huazhong University of Science and Technology, Wuhan, China, and the Ph.D. degree from City University of Hong Kong, Hong Kong SAR, China, all in electrical engineering in 2015, 2017, and 2021, respectively.
He is currently an Assistant Professor at the Robotics and Autonomous Systems Thrust, Systems Hub, The Hong Kong University of Science and Technology (Guangzhou), Guangzhou, China, and also an Affiliate Assistant Professor at the Department of Electronic and Computer Engineering, The Hong Kong University of Science and Technology, Hong Kong SAR, China. He was a Research Fellow at The University of Hong Kong, Hong Kong SAR, China, in 2021.
His research interests include electric machines and drives, applications of motor drives in robots, and electrified transportation.
\end{IEEEbiography}

\end{document}